\documentclass[%
 reprint,
bibnotes,
 amsmath,amssymb,
 aps
]{revtex4-1}

\usepackage{graphicx}
\usepackage{dcolumn}
\usepackage{bm}
\usepackage{hyperref}

\begin{document}


\title{Fundamental noise dynamics in cascaded-order Brillouin lasers}

\author{Ryan O. Behunin}
\affiliation{Department of Physics and Astronomy, Northern Arizona University, Flagstaff, AZ 86011}
\author{Nils T. Otterstrom}
\affiliation{Department of Applied Physics, Yale University, New Haven, CT 06520 USA.}
\author{Sarat Gundavarapu}
\affiliation{Department of Computer and Electrical Engineering, University of California at Santa Barbara, Santa Barbara, CA 93106 USA}
\author{Daniel J. Blumenthal}
\affiliation{Department of Computer and Electrical Engineering, University of California at Santa Barbara, Santa Barbara, CA 93106 USA}
\author{Peter T. Rakich}
\affiliation{Department of Applied Physics, Yale University, New Haven, CT 06520 USA.}

\date{\today}

\begin{abstract}
The dynamics of cascaded-order Brillouin lasers make them ideal for applications such as rotation sensing, highly coherent optical communications, and low-noise microwave signal synthesis. Remarkably, when implemented at the chip-scale, recent experimental studies have revealed that Brillouin lasers can operate in the fundamental linewidth regime where optomechanical and quantum noise sources dominate. 
To explore new opportunities for enhanced performance, we formulate a simple model to describe the physics of cascaded Brillouin lasers based on the coupled mode dynamics governed by electrostriction and the fluctuation-dissipation theorem. From this model, we obtain analytical formulas describing the steady state power evolution and accompanying noise properties, including expressions for phase noise, relative intensity noise and power spectra for beat notes of cascaded laser orders. Our analysis reveals that cascading modifies the dynamics of intermediate laser orders, yielding noise properties that differ from single-mode Brillouin lasers. These modifications lead to a Stokes order linewidth dependency on the coupled order dynamics and a broader linewidth than that predicted with previous single order theories. 
We also derive a simple analytical expression for the higher order beat notes that enables calculation of the Stokes linewidth based on only the relative measured powers between orders instead of absolute parameters, yielding a method to measure cascaded order linewidth as well as a prediction for sub-Hz operation. We validate our results using stochastic numerical simulations of the cascaded laser dynamics.
\end{abstract}
     
\maketitle

\section{Introduction}
Highly-coherent integrated photonic lasers will play an increasingly important role in a wide range of applications including low-noise microwave photonics \cite{Fortier2011}, atomic clocks \cite{Ludlow2015}, optical frequency synthesis, spectroscopy and rotation sensing \cite{Spencer2017, Rafac2000, Li2017,Liang2017}, coherent fiber communications \cite{Ip2008}, Doppler velocimetry \cite{Mocker1989} and high-resolution spectroscopy \cite{Hemmerich1990}. Photonic integration of these high-performance lasers is entering the era where it is feasible to implement chip-level functionalities that push sub-Hz linewidths, have low relative intensity noise (RIN), and have extremely low frequency jitter---performance typically requiring lab-based systems.  In spite of these impressive demonstrations, the theoretical description of these integrated lasers is not yet complete, and a full understanding of the complex steady-state and fast laser dynamics that determine the fundamental laser linewidth, RIN, center frequency jitter and technical noise is lacking \cite{Newbury2006}. 
With a more complete understanding of these dynamics, we can develop tools to measure and optimize the performance of these highly-coherent integrated lasers. 

Semiconductor laser emission linewidths in the range of 10 Hz to several-100 Hz are traditionally based on external cavity designs using discrete \cite{Morton2017} or hybrid-chip \cite{Liang2015,Fan2017} components in combination with frequency and phase locking feedback control. These designs make it possible to lower the fundamental laser linewidth, defined by a small number of terms given in the Schawlow-Townes linewidth \cite{Schawlow1958}, by combining techniques to increase the total number of photons in the cavity, decrease the cavity decay rate, and decrease the number of noise modes.

Another class of high-performance lasers utilizes stimulated Brillouin scattering (SBS). By leveraging unique dynamics that inhibit pump noise transfer \cite{Smith1991,Debut2001} and suppress RIN \cite{Stepien2002,Molin2008,Geng2007}, these lasers are capable of sub-Hz linewidth emission \cite{Smith1991}. Early fiber Brillouin lasers demonstrated $<$ 30 Hz intrinsic linewidth \cite{Smith1991} while Brillouin lasers utilizing externally coupled high-Q whispering gallery mode resonators (WGMR) \cite{Grudinin2009,Lee2012,Suh2017} achieve frequency noise indicative of sub-Hz intrinsic linewidths. Integration of Brillouin lasers onto a waveguide platform offers tremendous opportunities for reduced size, lower cost, and improved performance. Integrated Brillouin lasers have been created using a hybrid chalcogenide waveguide ring resonator bonded to a silicon photonic bus \cite{Morrison2017}, and in engineered photonic-phononic silicon waveguides \cite{Otterstrom2017}. However, at present, the properties of these lasers, with large Brillouin gain and relatively large optical losses, produce modest linewidths ($\sim$10-100 kHz). Recently, an integrated Brillouin laser based on a SiN waveguide with low Brillouin gain and low optical losses has been reported \cite{Gundavarapu2017a}. By harnessing these properties, this laser can produce sub-Hz Brillouin laser emission, bringing fiber-like performance to the chip-scale \cite{Gundavarapu2017a}. 

Cascaded-order Brillouin lasers are particularly suited to an array of technologies. These lasers produce multiple highly coherent emission lines that are spaced at microwave frequencies, making them ideal for microwave photonics, sensing, navigation, and timing applications. Cascading is produced when energy transfers from lower to higher order emission lines, a process that can be made more efficient by resonantly enhancing the Brillouin gain or by using additional optical gain mechanisms \cite{Lim1998, Zhan2006}. For example, compact Brillouin lasers using waveguide resonators, or discrete microresonators, are highly efficient at generating cascaded orders, due to high cavity Q in combination with optical and/or acoustic confinement \cite{Li2012,Suh2017,Gundavarapu2017a}. The emergence of these high-performance lasers, in addition to their value in applications, has created a pressing need for self-consistent models of noise dynamics in these multi-order Brillouin laser systems.

In this paper, we present the first theoretical investigation of lasing dynamics and noise properties of cascaded-order bulk, microcavity or photonic integrated Brillouin (or Raman) lasers. This investigation is based on a cascaded-order Brillouin laser model that builds on validated theories of single-mode Brillouin lasers and cascaded Raman lasers \cite{ Kippenberg2004,Vahala2008a,Li2012,Loh2015,Matsko2012,Otterstrom2017}. By assuming that the acoustic fields decay rapidly in comparison to the optical fields, we derive an approximate set of coupled nonlinear stochastic equations that describe the cascaded-order Brillouin laser dynamics.  These laser equations describe the energy transfer dynamics between the various optical modes and reveal rich noise dynamics generally described by colored multiplicative (spontaneous-spontaneous) processes. In agreement with prior work \cite{Kippenberg2004,Suh2017}, we find steady state energy exchange relations between various cascaded orders that reveal threshold and clamping behaviors as well as asymmetries for the even and odd Stokes orders,  with properties that are reminiscent of the behavior of Raman lasers. By linearizing about the steady-state for small amplitude, we find a simple compact set of equations describing the time-evolution of the phase and amplitude, and under the condition of perfect phase-matching, these linearized phase and amplitude dynamics decouple. These equations show that energy exchange between adjacent laser orders leads to complex relaxation oscillation dynamics, and besides reproducing the known results of single-mode Brillouin lasers \cite{Li2012,Loh2015,Matsko2012}, we find power spectra for cascaded order laser noise. Our model shows that cascading increases the noise of intermediate laser orders, broadening the linewidth by as much as a factor 3 and enhancing the RIN by as much as 30 dB at low frequencies. This enhancement occurs when cascaded orders inject spontaneous anti-Stokes photons into lower orders.

As an application of these phase dynamics, we calculate the phase noise for beat notes between neighboring laser orders, which to date has only been performed for pump-Stokes beat notes \cite{Matsko2012}. This result can be used to assess the coherence of microwave signals that are synthesized using cascaded Brillouin lasers. In addition, we show that measurements of the beat note phase noise and the relative powers of the participating optical fields enable precise linewidth measurements of the individual optical fields. Being insensitive to 
variations in component fabrication parameters and changes to these parameters as operating and environmental conditions change, this result can enable high resolution linewidth measurement of ultra-narrow linewidth lasers using heterodyne detection techniques.

This paper is structured as follows. 
In Sec. II, we describe the physics of cascaded Brillouin lasers. 
Sec. III describes the laser model, given in terms of a Hamiltonian describing the opto-acoustic interactions of a cascaded Brillouin laser system, from which the laser dynamics is described in terms of Heisenberg-Langevin equations that include quantum and thermal fluctuations. The equations of laser dynamics are simplified using adiabatic phonon field approximation and the steady state amplitude equations are analytically derived. Using this formalism in Sec. IV, we derive a simple set of analytical equations for threshold and clamped powers for a cascaded Brillouin laser system. In Sec. V, we formulate the amplitude and phase dynamics of individual optical modes and use these equations to find the power spectra describing RIN and phase noise. We derive the phase noise of a beat note between arbitrary Stokes orders and show how it can be used as an effective technique to characterize the noise properties of individual Stokes tones. We corroborate our amplitude and noise models using stochastic simulations of the Heisenberg-Langevin equations. Finally, we discuss the future directions to our work. 

In the next section we discuss the underlying physics that governs the dynamics and characteristics of cascaded Brillouin lasers. 

\section{Cascaded Brillouin Laser Physics}
\subsection{Brillouin coupling \& lasing}
\begin{figure}
\begin{center}
\includegraphics[width=3.4in]{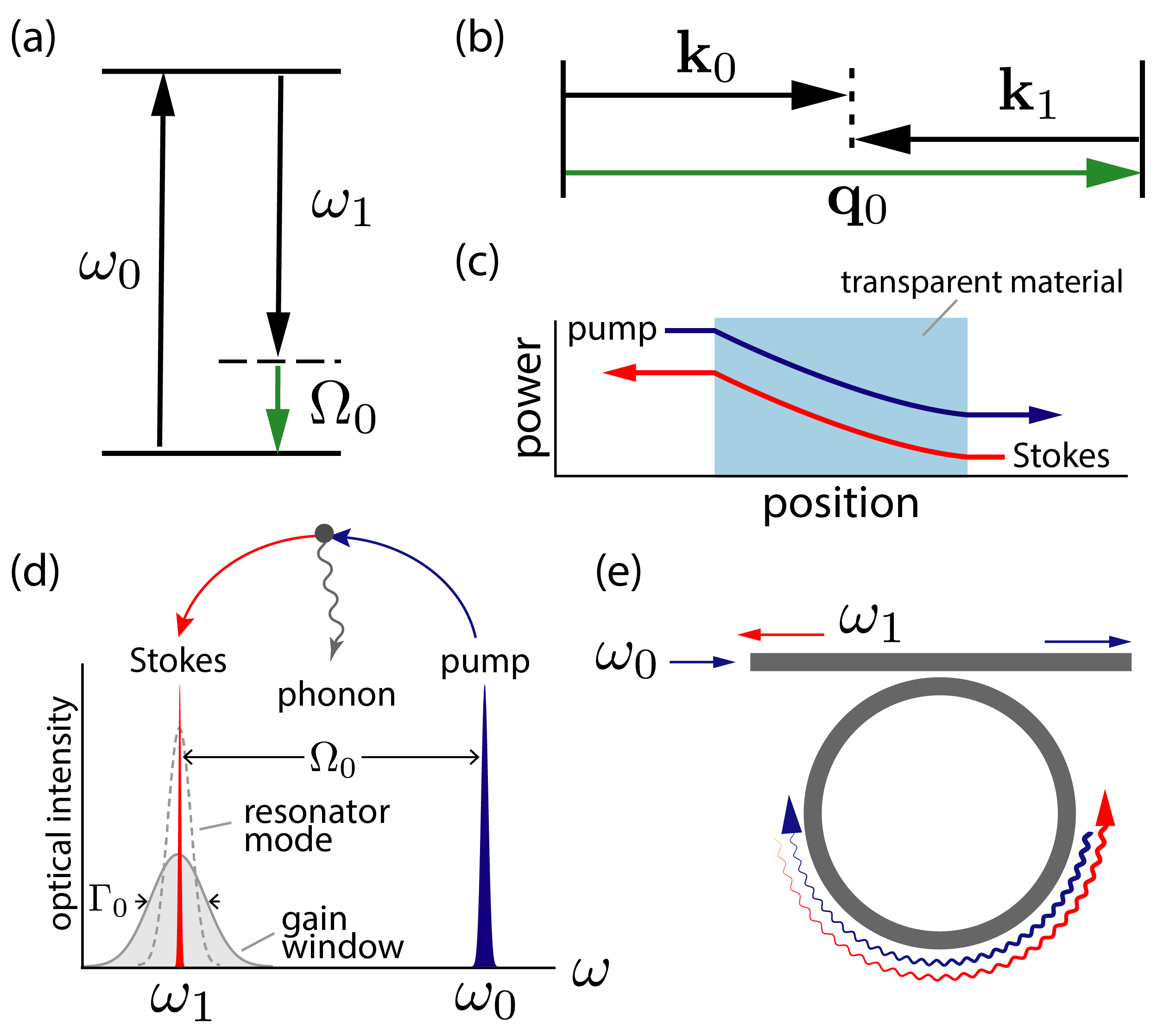}
\caption{Fundamentals of Brillouin lasing. (a) Energy conservation and (b) wavevector phase matching requirements. Brillouin coupling mediated (c) optical amplifier and (d) energy transfer. (e) Ring resonator-based Brillouin laser.}
\label{intro-fig}
\end{center}
\end{figure}

Brillouin coupling, enabling light scattering from traveling sound waves, is the key physics permitting Brillouin lasing \cite{Boyd2003}. 
By optically pumping a transparent medium, Brillouin coupling can be used to create an optical amplifier (see Fig. \ref{intro-fig}c). 
Through this nonlinear optomechanical process, a high-frequency (pump) photon of frequency $\omega_{0}$ and wavevector ${\bf k}_{0}$, can decay into a lower frequency (Stokes) photon and a phonon with respective frequencies $\omega_{1}, \Omega_0$ and wavevectors ${\bf k}_{1},{\bf q}_0$ Fig. \ref{intro-fig}a \& b.
Provided that phase-matching is satisfied, i.e. $\omega_{0} = \omega_{1} +\Omega_0$ (energy conservation) and ${\bf k}_{0} = {\bf k}_{1} +{\bf q}_0$ (akin to momentum conservation), Brillouin coupling can efficiently transfer energy from the pump mode to the Stokes mode. 
Similar to gain media for laser systems with inverted populations, a Stokes photon can stimulate the decay of a pump photon into a Stokes photon, thereby producing stimulated emission and optical amplification. 
This amplification process occurs within a narrow gain window, at frequency determined by the phase matching conditions, and with a width given by the decay rate $\Gamma_0$ of the participating phonons Fig. \ref{intro-fig}d. 
For backward Brillouin scattering, where the Stokes wave propagates antiparallel to the copropagating pump and phonon, phase matching places the gain window at 
$
\omega_{1} \approx \left(1 - (2nv/c)\right) \omega_0
$
where $n$ and $v$ are the material's index of refraction and longitudinal sound speed and $c$ is the speed of light \cite{Boyd2003}. For silica glass, $(2nv/c) \sim 5 \times 10^{-5}$, making $ \omega_{0} - \omega_{1} \sim (2\pi)11$ GHz for a pump wavelength of 1.5 $\mu$m. 
By pumping an optical resonator that supports an optical mode at the Stokes frequency, the Brillouin gain window overlaps with $\omega_{1}$, and a Brillouin laser can be created (see Fig. \ref{intro-fig}e). 

\subsection{Cascaded Brillouin lasing}
\begin{figure}
\begin{center}
\includegraphics[width=3.3in]{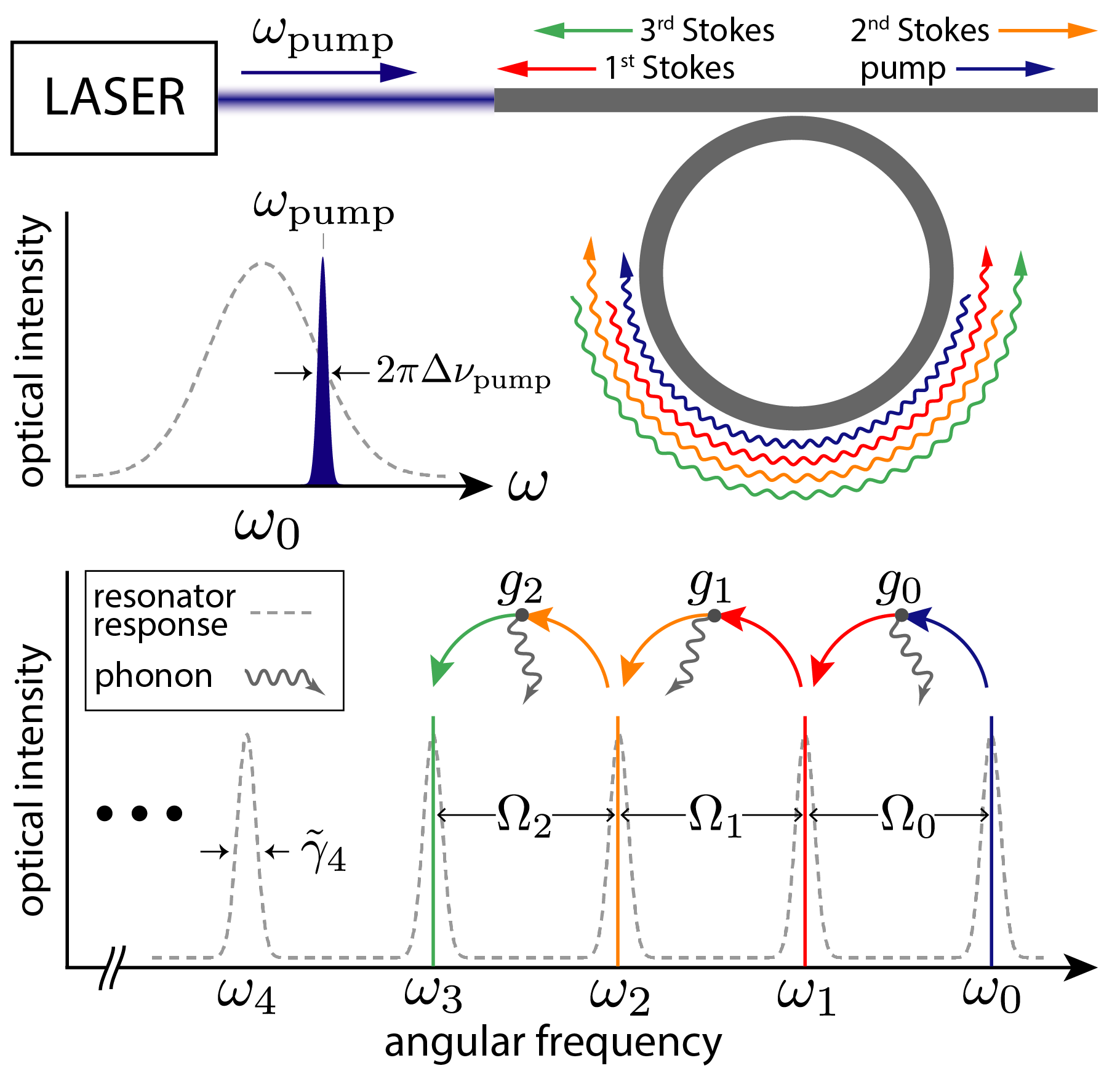}
\caption{Illustration of a cascaded Brillouin laser. A laser of frequency $\omega_{\rm pump}$ and linewidth $\Delta\nu_{\rm pump}$ pumps an optical resonator. Light in the $\omega_0$-mode (blue) can scatter to $\omega_1$ (red) by emitting a phonon. When lasing the $\omega_1$-optical mode can act as a pump for higher Stokes orders. Here, $g_m$ and $\tilde{\gamma}_m$ respectively quantify the Brillouin coupling rate between the $m$ and the $(m+1)$th modes and the optical decay rate of the $m$th mode.}
\label{cascaded-laser-sketch}
\end{center}
\end{figure}

At high Stokes laser intensities, cascaded Brillouin lasing can occur (see Fig. \ref{cascaded-laser-sketch}). Under this condition, the Stokes field (red of Fig. \ref{cascaded-laser-sketch}) acts as a pump for a counterpropagating second Stokes (orange) order with frequency $\omega_2$. This process is mediated by a distinct phonon, with frequency $\Omega_1$ and propagating in the opposite direction as the phonon participating in the pump to Stokes energy transfer. Consequently, the pump-Stokes frequency difference is roughly $\sim(2\pi)$600 kHz greater than the Stokes-Stokes 2 frequency difference in silica and for a pump wave length of 1.5 $\mu$m. In high quality factor resonators with evenly spaced modes, this frequency shift can produce walk-off that can stifle further cascading. However, provided that the resonator supports an optical mode near $\omega_2$ (or any successive order), i.e. within the gain window, cascaded lasing of the second, or higher order, Stokes mode(s) can be produced. With sufficiently high pump powers and given a resonator supporting optical modes at higher order Stokes frequencies, cascading can continue to many orders, each cascaded order pumped by the previous order and mediated by a distinct phonon. 

This cascaded lasing behavior naturally occurs in WGMR and ring resonators, where the optical modes are regularly spaced by the cavity free-spectral range (FSR) (see Fig. \ref{intro-fig}(e)). For a typical system, the gain bandwidth and optical cavity linewidths are much larger than the walk-off produced by successive phase-matching as described above, and consequently these systems can produce cascaded lasing to many Stokes orders \cite{Lee2012,Li2013,Suh2017,Gundavarapu2017a}. As concrete examples, we base the laser modeling to follow on integrated waveguide-based Brillouin lasers of the type described above.  

\subsection{Brillouin laser noise}   
\begin{figure}
\begin{center}
\includegraphics[width=3.4in]{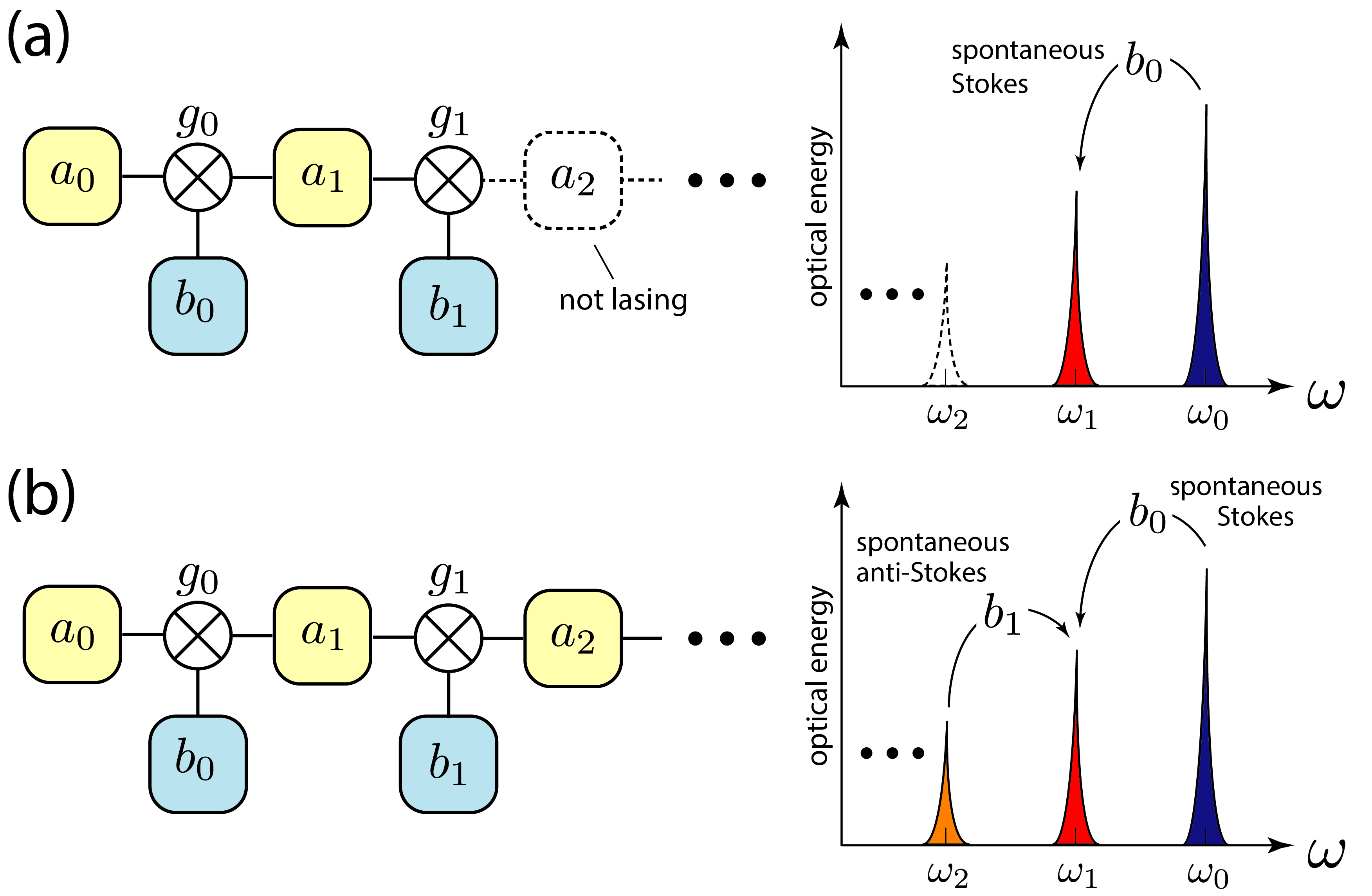}
\caption{Illustration of noise dynamics in cascaded Brillouin lasers. Tiles represent optical and acoustic modes. The mixer symbol represents the nonlinear optomechanical coupling between two optical and one acoustic mode. (a) Below threshold for cascaded lasing, optomechanical coupling enables noise transfer between the $m=0$ and the $m=1$ through spontaneous Brillouin scattering from the phonon mode $b_0$. (b) Above threshold for cascaded lasing, noise can be injected into the $m=1$ mode from spontaneous scattering from thermal phonons in the $b_0$ and $b_1$ modes.}
\label{summaryFig}
\end{center}
\end{figure}

We show that the noise dynamics of Brillouin lasers are distinct above and below the threshold for cascaded Brillouin lasing. This is because cascaded lasing opens new noise channels that are absent in uncascaded Brillouin lasers. We explain the origin of this behavior in Fig. \ref{summaryFig}, which considers energy transfer dynamics to and from an optical mode $a_1$.
 
Optomechanical coupling produces a nonlinear interaction between three waves in a manner that is similar to a mixer (see Fig. \ref{summaryFig}), where the mixer output frequency is given by the sum and difference of the two injected tones. Using this analogy, we can explain the optomechanical noise present in Brillouin lasers. For example, when a coherent field in the optical mode $a_0$ and a noisy acoustic field $b_0$ (due to thermal fluctuations) are injected into neighboring mixer ports, the mixer output comprises of a coherent carrier with two noisy sidebands. In a Brillouin laser, the frequency of the lower sideband is given by $\omega_1$, and as a result this spontaneous Stokes scattering process injects noise into the $a_1$ mode. Likewise, a coherent field present in the optical mode $a_1$ can also mix with an acoustic field to produce a carrier with noisy sidebands. However, in this case the frequency of the higher sideband is given by $\omega_0$, thereby transferring noise from the acoustic field to the $a_0$ mode through spontaneous anti-Stokes scattering (see Fig. \ref{summaryFig}(a)).

Below the cascaded lasing threshold, the optical mode $a_2$ is neither coherent nor occupied with a large number of quanta. In other words, $a_2$ is noisy and fluctuates in amplitude around zero. While this noisy field, in addition to a noisy acoustic field $b_1$, can be injected into the two ports of a mixer to produce multiplicative (spontaneous-spontaneous) noise in the $a_1$ mode, the magnitude of this noise source is small because the thermal occupation (quantifying the noise amplitude) of the optical mode $a_2$ is essentially zero. 

However, once cascaded lasing is achieved, the coherent field now present in the $a_2$ mode can efficiently transfer noise from the acoustic mode $b_1$ to $a_1$ (as seen in Fig. \ref{summaryFig}(b)), coupling the optical mode $a_1$ to an additional heat bath. 
We find that these new noise channels, introduced by cascading, enhance the phase and amplitude noise, thereby producing contrasting behaviors from single-mode Brillouin lasers \cite{Debut2000,Li2012,Loh2015}. For a lasing order with a fixed emitted power, we find that presence of additional laser orders (due to cascading) can alter the laser linewidth by as much as a factor 3, and enhance the relative intensity noise by nearly 30 dB at low frequencies. 

\section{Theory}
{\it Model Hamiltonian:}
The physics of a cascaded Brillouin laser can be described by the model Hamiltonian $H$ given by 
\begin{align}
\label{Hamiltonian}
H =  &\hbar   
\sum_m [\omega_{ m} a^\dag_{m} a_{m} + \Omega_m b_m^\dag b_m
\nonumber \\
& \quad\quad \quad\quad+ (g_{m}  a^\dag_{m} a_{m+1} b_m + {\rm H.c.})],
\end{align}
and schematically represented in Figs. \ref{cascaded-laser-sketch} \& \ref{summaryFig}.
This model generalizes the treatment of optomechanical laser noise described in prior work \cite{Vahala2008a,Li2012,Matsko2012,Loh2015,Otterstrom2017} to include the effects of cascaded lasing.   
Here, $a_{m}$ and $b_m$ are the respective annihilation operators for the $m$th optical and phonon modes, with respective frequencies $\omega_m$ and $\Omega_m$. The mode index $m$ labels the cascaded Stokes order, $m=0$ corresponding with the pump, $m=1$ corresponding with the first Stokes order, etc. 
In contrast with linear waveguides, where mode amplitudes can change along the system's symmetry direction, our model treats the field within the optical and acoustic resonator as independent of space, and essentially composed of a pure $k$-vector mode (either traveling or standing), this aspect of our model contrasts with the work of Debut {\it et al.} \cite{Debut2000} which accounts for the spatial dynamics of the optical field throughout the laser resonator. This approximation is valid so long as the loaded optical decay rate is much smaller than the free spectral range of the resonator. The coupling rate $g_m$ quantifies the Brillouin interaction between $m$th phonon mode and the $m$th and $(m+1)$th optical modes, including the effects of spatial phase matching. This coupling rate, determined by the spatial overlap of the acoustic and optical modes, is discussed in detail in Appendix A.  

{\it Heisenberg-Langevin equations:} The laser dynamics are described by the Heisenberg-Langevin equations of motion resulting from Eq. \eqref{Hamiltonian}. In a frame rotating at the resonance frequency of each field, we find
\begin{align}
\label{Heisenberg-Langevin-1}
 \dot{a}_{m}  = &-\frac{1}{2}\gamma_{m}a_{m} + \sqrt{\gamma_{\rm ext}} F_{\rm pump} e^{-i\omega_m t}\delta_{m0} +\eta_{m} 
\\ \nonumber 
& 
 -i g_m a_{m+1} b_m e^{-i\Delta \omega_{m+1} t}-i g^*_{m-1} b^\dag_{m-1} a_{m-1} e^{i\Delta \omega_{m} t}  \\
 \label{Heisenberg-Langevin-2}
 \dot{b}_m = &  - \frac{1}{2}\Gamma_m b_m + \xi_m
 -i g^*_m a_{m+1}^\dag a_{m} e^{-i\Delta \omega_{m+1} t} .
\end{align}
Here, we have added the phenomenological decay rates, $\gamma_{m}$ and $\Gamma_m$, for the respective $m$th optical and acoustic modes, and the Langevin forces $\eta_m$ and $\xi_m$ to equations of motion by hand. These terms describe the noise and dissipation present in each degree of freedom. We require that these terms yield a state of thermal equilibrium in the absence of electrostrictive coupling. Above, the parameter $\Delta \omega_m$ is the difference in resonance frequencies given by $\omega_m -\omega_{m-1} + \Omega_{m-1}$, and $\gamma_{\rm ext}$ denotes the component of the optical mode decay rate due to coupling of the laser resonator to a bus waveguide that supplies power to the laser. The time-dependent function $F_{\rm pump}$, representing the optical pump, supplies power to the $m=0$ mode of the resonator. This function is normalized such that $|F_{\rm pump}|^2$ is given in units of photon flux, so that the power supplied to the laser through the bus waveguide $P_{\rm pump}$ is given by $\hbar \omega_{\rm pump} |F_{\rm pump}|^2$. In addition, we assume that the noise of this source laser is dominated by phase noise, as a result we assume that $P_{\rm pump}$ is time-independent and all of the time-dependence of $F_{\rm pump}$ is described in terms of a random time-dependent phase $\varphi_{\rm pump}$. This randomly varying phase models a pump laser with a finite linewidth (see Fig. \ref{cascaded-laser-sketch}). The Langevin forces $\eta_{m}$ and $\xi_m$ quantify the quantum and thermal fluctuation of the respective optical and acoustic fields. These Langevin forces are zero-mean Gaussian random variables with white power spectra \cite{Vahala2008a,Li2012,Loh2015}, yielding the correlation properties given by 
\begin{align}
\label{Langevin-corr}
&\langle \eta_m^\dag(t) \eta_{m'}(t') \rangle = \gamma_{m} N_m \delta(t-t') \delta_{mm'} \\
&\langle \eta_m(t) \eta^\dag_{m'}(t') \rangle = \gamma_{m} (N_m+1) \delta(t-t') \delta_{mm'} \\
&\langle \xi_m^\dag(t) \xi_{m'}(t') \rangle = \Gamma_m n_m \delta(t-t') \delta_{mm'} \\
&\langle \xi_m(t) \xi_{m'}^\dag(t') \rangle = \Gamma_m (n_m+1) \delta(t-t') \delta_{mm'}.
\end{align}
where $N_{m}$ and $n_{m}$ are the thermal occupation numbers of the $m$th optical and acoustic modes (i.e. $N_{m} = (\exp\{\hbar \omega_m/k_B T \} -1)^{-1}$ and $n_{m} = (\exp\{\hbar \Omega_m/k_B T \} -1)^{-1}$) and $\langle ... \rangle$ denotes an ensemble average with respect to the Langevin forces. 

{\it Adiabatic elimination of phonon fields:}
In many Brillouin lasers the decay rate of the relevant acoustic modes is much larger than the decay rate of the participating optical modes (i.e. $\Gamma_m  \gg \gamma_m \ {\rm and} \ \gamma_{m+1}$). For near-resonant systems (i.e. $\Delta \omega_m \ll \Gamma$) possessing this separation of time-scales, the phonon fields adiabatically follow the electrostrictive forcing generated by the beat notes of the various optical fields. In this limit, we find the approximate solution for the phonon dynamics given by 
\begin{align}
b_m &\approx - ig_m^* \chi_m a^\dag_{m+1} a_m + \hat{b}_m,
\end{align}
where $\chi_m \equiv (-i\Delta\omega_{m+1}+\Gamma_m/2)^{-1}$ and $\hat{b}_m$, quantifying the thermal and quantum fluctuations of the phonon field, is given by 
\begin{align}
\label{b_Langevin}
\hat{b}_m = \int^t_{-\infty} d\tau e^{- \frac{\Gamma_m}{2}(t-\tau)}\xi_m(\tau). 
\end{align}
Physically, this approximation is valid so long as the electrostrictive forces change so slowly in time that the phonon assumes its steady-state amplitude at each instant. 

By adiabatically eliminating the phonon field, we can obtain a simplified set of equations describing the dynamics of a cascaded Brillouin laser. By combining the approximate solution for $b_m$ with Eqs. \eqref{Heisenberg-Langevin-1} \& \eqref{Heisenberg-Langevin-2}, we find the effective equation of motion for the optical field amplitudes given by
\begin{align}
\label{laser_equation_1}
\dot{a}_m = & -\left( \frac{\tilde{\gamma}_{m}}{2}  + \mu_m a_{m+1}^\dag a_{m+1} - 
\mu^*_{m-1} a_{m-1}^\dag a_{m-1} \right) a_{m} \nonumber \\
&   + h_{m} + \sqrt{\gamma_{\rm ext}} F_{\rm pump} e^{-i\omega_m t}\delta_{m0},
\end{align}
where $\mu_m \equiv |g_m|^2 \chi_m$ and $\tilde{\gamma}_m \equiv \gamma_m  + 2 \mu_m$, and the Langevin force $h_{m}$, defined by $h_{m} = \eta_{m} - i g_m a_{m+1} \hat{b}_m - i g^*_{m-1}\hat{b}^\dag_{m-1} a_{m-1}$, describes the colored multiplicative noise imparted to the optical fields through electrostrictive coupling in addition to quantum and thermal fluctuations of the optical modes. 
The function $\mu_m$ 
is the nonlinear susceptibility associated with Brillouin scattering, $2{\rm Re}[\mu_m]$ yielding the Brillouin amplification rate per photon. This function can be related to the Brillouin gain factor $G_{{\rm B},m}$, quantifying the spatial rate of Brillouin-mediated energy transfer per Watt of pump power along a waveguide, through the relation 
\begin{align}
\label{}
  G_{{\rm B},m} = \frac{2{\rm Re}[\mu_m] L}{\hbar \omega_{m-1} v_{g,m}v_{g,m-1}}
\end{align}
where $L$ is the resonator length and $v_{g,m}$ is the group velocity of the $m$th optical mode. This gain factor $G_{{\rm B},m}$ can also be expressed in terms of the bulk gain $g_{{\rm B},m}$, in units of meter per Watt, by multiplying by the effective acousto-optic overlap area $A_{\rm eff}$, i.e. $g_{{\rm B},m} = A_{\rm eff} G_{{\rm B},m}$. Through energy transfer measurements of Brillouin scattering in a waveguide segment with the same properties as waveguide used to create the laser resonator, the gain factor $G_{{\rm B},m}$ can be obtained and the electrostrictive coupling rate $g_m$ can be derived. 

Equation \eqref{laser_equation_1} shows that the dynamics of the $m$th mode exhibit threshold behavior. Namely, when 
$
{\rm Re}[\mu^*_{m-1}] a_{m-1}^\dag a_{m-1} >  \frac{\tilde{\gamma}_{m}}{2}  + {\rm Re}[\mu_m] a_{m+1}^\dag a_{m+1}
$
the $m$th mode becomes unstable, and the mode amplitude can grow in magnitude. When this laser threshold condition is satisfied, a small fluctuation of $a_m$ can be be amplified, creating strong coherent laser oscillation, so that $a_m$ acquires a nonvanishing mean value. This behavior is analogous to a second order phase transition \cite{DeGiorgio1970}, where the average amplitude plays the role of the order parameter. 

Above laser threshold, it is convenient to describe the dynamics of $a_m$ using the following decomposition
\begin{equation}
\label{laser-ansatz}
a_m = (\alpha_m + \delta \alpha_m)e^{i\varphi_m}.
\end{equation} 
Here, $\alpha_m$ is the time-independent steady-state laser amplitude, and $\delta \alpha_m$ and $ \varphi_m$ are time-dependent fluctuations of the respective amplitude and phase of the laser emission from the $m$th optical mode. These zero-mean fluctuating quantities describe the laser noise properties, $\delta \alpha_m$ describing the relative intensity noise (RIN), and $ \varphi_m$ the phase noise. In the following sections, we use the representation of $a_m$ above to derive the steady-state laser dynamics of the optical modes and to describe the fundamental noise properties of cascaded Brillouin lasers. 

\section{Steady-state laser amplitudes, threshold, and cascading}
By inserting the representation of $a_m$ given by Eq. \eqref{laser-ansatz} into Eq. \eqref{laser_equation_1}, and taking the time average, we find the following recursion relation between the various laser amplitudes and the time-independent mean amplitude of the pump field $|F_{\rm pump}|$
\begin{align}
\label{recursion}
\left( \frac{\tilde{\gamma}_{m}}{2}  + \mu'_m \alpha^2_{m+1} - 
\mu'_{m-1} \alpha^2_{m-1} \right) \alpha_{m} =  \sqrt{\gamma_{\rm ext}}|F_{\rm pump}| \delta_{m0}.
\end{align}
Here, $\mu'_m = {\rm Re}[\mu_m]$ and we have dropped nonvanishing terms of order $\delta\alpha_m^2$ (away from threshold $\alpha_m \gg |\delta \alpha_m|$). Equivalently, this recursion relation can be written in terms of the coherent occupation numbers $p_m = \alpha_m^2$ yielding the steady-state equations for the mode occupation number given by
\begin{align}
\label{}
\left(\!\frac{\tilde{\gamma}_{m}}{2}\!+\!\mu'_m p_{m+1}\!-\! 
\mu'_{m-1} p_{m-1} \!\right) p_{m}  = \sqrt{\gamma_{\rm ext}}|F_{\rm pump}| \sqrt{p_m} \delta_{m0}.
\end{align}
These recursion relations are analogous to prior results for cascaded Raman lasers (see steady-state limit of Eq. 10 in Ref. \cite{Kippenberg2004}), which can be modeled by the Hamiltonian given in Eq. \eqref{Hamiltonian}. 

We can use these recursion relations to find the emitted laser power for each mode. To obtain the emitted power of the $m$th mode $P_m$, one first obtains the intracavity power by multiplying the occupation number $p_m$ by the energy stored in the resonator per photon $\hbar \omega_m v_{g,m}/L$, where $L$ is the length of the resonator and $v_{g,m}$ is the group velocity of the $m$th mode. By multiplying the intracavity laser power by resonator-bus waveguide power coupling factor $\kappa$, we obtain $P_m$ given by 
\begin{align}
\label{power}
 P_m = \frac{\hbar \omega_m v_{g,m}\kappa}{L} p_m. 
\end{align}

Now we describe how the recursion relation Eq. \eqref{recursion} can be used to find the steady-state laser powers. When only $k$ orders are lasing, we know $\alpha_{k+1} = 0$, and given that the anti-Stokes mode to the pump cannot lase we know $\alpha_{-1} = 0$. These two conditions can be used with Eq. \eqref{recursion} to give 
\begin{align}
\label{01-relation}
& \left( \frac{\tilde{\gamma}_{0}}{2}  + \mu'_0 \alpha^2_{1}  \right) \alpha_{0} =  \sqrt{\gamma_{\rm ext}}|F_{\rm pump}|  \quad {\rm for} \quad m=0
\\
\label{threshold}
&\alpha^2_{k-1} =   \frac{\tilde{\gamma}_{k}}{2 \mu'_{k-1} }
\\
\label{recursion-2}
&\alpha^2_{m-1} =  \frac{\mu'_{m}}{\mu'_{m-1}} \alpha^2_{m+1} + \frac{\tilde{\gamma}_{m}}{2 \mu'_{m-1}}
 \quad {\rm for} \quad k>m>0.
\end{align}
There are a number of important results that can be drawn from these equations. First, for $k$th order cascading, the $k-1$ mode is clamped. As a result, the recursion relation Eq. \eqref{recursion-2} implies that the $k-3,k-5,k-7.....$ are clamped as well. This behavior is illustrated in Fig. \ref{powerFig} which shows the emitted powers of each Stokes order as function of power supplied to the laser $P_{\rm pump}$. In other words, if $k$ is even, all odd orders are clamped, and if $k$ is odd all even orders are clamped. 

Combing the results of Eqs. \eqref{01-relation}, \eqref{threshold}, and \eqref{recursion-2}, the power in the $2m$th and the $(2m+1)$th orders are respectively given in terms of the $\alpha_0$ and $\alpha_1$ as 
\begin{align}
\label{ss-pow1}
 \alpha_{2m}^2   & = C^{(e)}_m ( \alpha_0^2  - S^{(e)}_m )
 \\
     \alpha_{2m+1}^2   & = C^{(o)}_m (\alpha_1^2  - S^{(o)}_m )
\end{align} 
where we the recursion formulas for the steady-state amplitudes yield 
\begin{align}
\label{ss-pow2}
C_m^{(e)} =  & \prod_{j=1}^m \frac{\mu'_{2j-2}}{\mu'_{2j-1}} 
\\
S_m^{(e)} = & \frac{1}{\mu'_0} \sum_{j=1}^m \frac{1}{2} \tilde{\gamma}_{2j-1} C^{(o)}_{j-1} 
 \\
C_m^{(o)} = &  \prod_{j=1}^m \frac{\mu'_{2j-1}}{\mu'_{2j}} 
\\
S^{(o)}_m = & \frac{1}{\mu'_0} \sum_{j=1}^m \frac{1}{2} \tilde{\gamma}_{2j} C^{(e)}_j. 
\end{align}
 
\subsection{Laser thresholds and powers}
Using the analysis given above, we obtain the power emitted from each mode and threshold powers for each order of cascaded lasing. We find the expressions for the emitted power, in terms of $P_0$ and $P_1$, given by 

\begin{align}
\label{ss-pow3}
 P_{2m}       & = C^{(e)}_m \frac{\omega_{2m} v_{g,2m}}{\omega_{0} v_{g,0}} \bigg[  P_0  - \frac{\hbar \omega_{0} v_{g,0}\kappa}{L} S^{(e)}_m \bigg]
 \\
 P_{2m+1}   & = C^{(o)}_m  \frac{\omega_{2m+1} v_{g,2m+1}}{\omega_{1} v_{g,1}} \bigg[  P_1  - \frac{\hbar \omega_{1} v_{g,1}\kappa}{L}S^{(o)}_m \bigg]
\end{align} 
which can be used to calculate the power emitted from any mode in terms of $P_0$ and $P_1$. 
To find $P_0$ and $P_1$, we must separately consider the cases when an even and odd number of Stokes orders are lasing. 

\subsection{Cascading to $2k+1$ orders (odd number of Stokes orders)}
First, we consider the case when an odd number of Stokes order are lasing. In this case, the powers of all even orders are clamped.  Using Eqs. \eqref{ss-pow3}, \eqref{01-relation} and \eqref{power}, we find 
\begin{align}
\label{}
  P_0 = &  \frac{\hbar \omega_0 v_{g,0}\kappa}{L} S^{(e)}_{k+1}     \\
  P_1 = &  \frac{\hbar \omega_1 v_{g,1}\kappa}{L} \frac{1}{\mu'_0}
  \left( \sqrt{ \frac{\gamma_{\rm ext}P_{\rm pump}}{\hbar \omega_{\rm pump}S^{(e)}_{k+1}}} -  \frac{\tilde{\gamma}_0}{2}\right).  
\end{align}

\subsection{Cascading to $2k$ orders (even number of Stokes orders)}
 In contrast, for an even number of Stokes orders all odd orders are clamped. Again, using Eqs. \eqref{ss-pow3}, \eqref{01-relation} and \eqref{power}, we find
 \begin{align}
\label{}
   P_0 =  &  
     \frac{ \omega_0 v_{g,0}\kappa}{L \omega_{\rm pump}} \left( \frac{\tilde{\gamma}_0}{2}+\mu'_0 S^{(o)}_k
     \right)^{-2}  \gamma_{\rm ext}P_{\rm pump} \\
   P_1 = & \frac{\hbar \omega_1 v_{g,1}\kappa}{L} S^{(o)}_k.   
\end{align}
Using the relations above we can determine the threshold power for cascading at an arbitrary order. Threshold for the $k$th order is met when the power in the $(k-1)$th becomes clamped. Using Eqs. \eqref{threshold} and \eqref{power}, this clamped power is given by 
\begin{align}
\label{}
P_{k-1} =  \frac{\hbar \omega_{k-1} v_{g,k-1}\kappa}{L} \frac{\tilde{\gamma}_{k}}{2 \mu'_{k-1}}
\end{align}
yielding the threshold power for the $k$th mode $P^{\rm th}_k$ given by
\begin{align}
\label{threshold-power}
 P^{\rm th}_{k} = \frac{\hbar \omega_{\rm pump}}{\gamma_{\rm ext}} \left\{
        \begin{array}{ll}
             S^{(e)}_{k/2}(\mu'_0 S^{(e)}_{k/2} + \tilde{\gamma}_0/2)^2 & \ k \ {\rm even} \\
             S^{(e)}_{(k+1)/2}(\mu'_0 S^{(o)}_{(k-1)/2} + \tilde{\gamma}_0/2)^2  
             & \  k \ {\rm odd}.
            \end{array}
             \right.
\end{align}

\subsection{Special case: $\tilde{\gamma}_m = \tilde{\gamma}$, $\mu_m = \mu$, and $v_{g,m} = v_g$}
Up to this point, we have accounted for the possibility that the gain and loss properties of the resonator may vary mode by mode. However, in many Brillouin laser resonators these properties are approximately constant over a large frequency range. In this short section we explore the steady-state laser physics for the case where $\tilde{\gamma}_m = \tilde{\gamma}$, $\mu_m = \mu$, and $v_{g,m} = v_g$, yielding a dramatic simplification of the analysis. Under these conditions $C^{(e)}_m = C^{(o)}_m = 1$ and $S^{(e)}_m = S^{(o)}_m = (\tilde{\gamma}/2\mu')m$, leading to the emitted powers given by
\begin{align}
\label{ss-pow4}
 P_{2m}       & = \omega_{2m} \bigg[\frac{P_0}{\omega_0}  - \frac{ \hbar \tilde{\gamma} \gamma_{\rm ext}}{2\mu'}  m \bigg]
 \\
 P_{2m+1}   & =  \omega_{2m+1}\bigg[\frac{P_1}{\omega_1}  - \frac{\hbar  \tilde{\gamma} \gamma_{\rm ext}}{2\mu' }m  \bigg],
\end{align} 
and the threshold powers given by 
\begin{align}
\label{}
  P^{\rm th}_{j}
  =  \frac{\hbar \omega_{\rm pump} \tilde{\gamma}^3}{64 \mu \gamma_{\rm ext}} \left\{
        \begin{array}{ll}
            j(j+2)^2 & \quad j \quad {\rm even} \\
            (j+1)^3   & \quad j \quad {\rm odd},
        \end{array}
    \right.
    \end{align}
where we have used $\gamma_{\rm ext} = v_g \kappa/L$. Next, we find the power emitted by each order.

\subsubsection{Cascaded lasing of $2k+1$ orders only}
When an odd number $2k+1$ of cascaded orders are lasing, we find the following expressions for the laser power emitted by the even $2m$ and odd $2m+1$ orders   
\begin{align}
\label{pwrDyn-1}
& P_{2m} = \frac{\hbar \omega_{2m} \gamma_{\rm ext} \tilde{\gamma}}{2\mu}(k+1-m)
 \\
 &P_{2m+1} =  \frac{4 \omega_{2m+1} \gamma_{\rm ext}^2}{
 \omega_{\rm pump} \tilde{\gamma}^2 }\frac{m\!+\!1}{(k\!+\!1)^3} \bigg[
 \sqrt{P^{\rm th}_{2k+1} \! P_{\rm pump}}\frac{k\!+\!1}{m\!+\!1}\! - \! P^{\rm th}_{2k+1}\bigg].
 \nonumber 
\end{align}

\subsubsection{Cascaded lasing of $2k$ orders only}
When an even number $2k$ of Stokes orders are lasing the emitted powers given by
\begin{align}
\label{pwrDyn-2}
 P_{2m}  & =  \frac{4 \omega_{2m} \gamma_{\rm ext}^2}{
 \omega_{\rm pump} \tilde{\gamma}^2 }\frac{1}{(k+1)^2} \bigg[
 P_{\rm pump}- \frac{m}{k} P^{\rm th}_{2k}\bigg]
 \\
 P_{2m+1} & =  \frac{\hbar \omega_{2m+1} \gamma_{\rm ext} \tilde{\gamma}}{2\mu}(k-m).  \nonumber
\end{align} 
Under the appropriate assumptions, these formulas reproduce the results of previous works on cascaded lasers \cite{Kippenberg2004}. 

\begin{figure}[h]
\begin{center}
\includegraphics[width=3.4in]{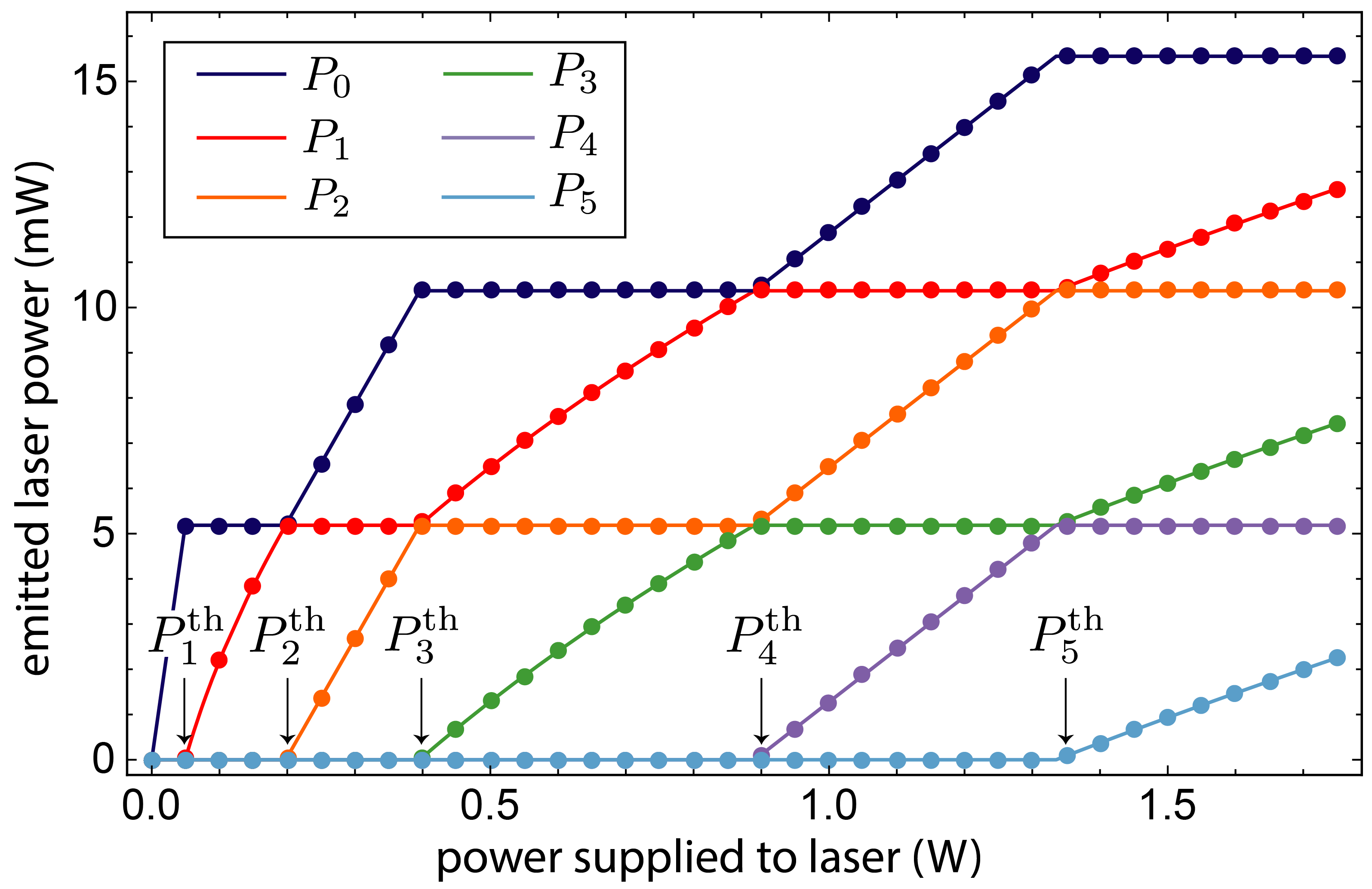}
\caption{Steady-state laser power for the laser parameters given in Tab. \eqref{Parameter-Table}. Solid lines represent theoretical predicitions for the steady-state powers given by Eqs. \eqref{ss-pow1} and \eqref{ss-pow2}, and open circles denote the steady-state powers obtained from stochastic simulations of Eq. \eqref{laser_equation_1}.} 
\label{powerFig}
\end{center}
\end{figure}

In Fig. \ref{powerFig}, we plot the emitted laser powers described by Eqs. \eqref{pwrDyn-1} and \eqref{pwrDyn-2}, and compare with the emitted powers obtained through stochastic simulations of Eqs. \eqref{Heisenberg-Langevin-1} \& \eqref{Heisenberg-Langevin-2}. The results displayed in Fig. \ref{powerFig} show that these analytical expressions accurately capture the steady-state laser dynamics.     

\begin{table}[]
 \caption{Cascaded Brillouin laser simulation parameters. The coupling rate $g$, optical decay rate $\tilde{\gamma}$, and the acoustic decay rate $\Gamma$ are the same for all considered orders.}
\label{Parameter-Table}
  \begin{center}
    \label{tab:table1}
    \begin{tabular}{|c|c|c|} 
      \hline
      $g$ 					& 1.54 kHz 		 & electrostictive coupling rate	\\
      $\Gamma$ 				& $(2\pi)$200 MHz	 & phonon decay rate	\\
      $\tilde{\gamma}$ 				& $(2\pi)$6.88 MHz	 & optical decay rate	\\
      $\omega_{\rm pump}$ 		& $(2\pi)$195.3 THz   & pump laser frequency	\\
      $G_{\rm B}$   			& 0.1(Wm)$^{-1}$       & Brillouin gain     \\
      $\kappa$   				& 0.0025 			  & power coupling    	\\
      $L$   					& 0.0743 m 		  & resonator length	 \\
      $v_g$   				& 2.08$\times10^8$ m/s & optical group velocity 	 \\
      $\gamma_{\rm ext}$   		& $(2\pi)$1.11 MHz 	   & external optical loss rate    \\
      $\Delta\nu_{\rm pump}$		& $(2\pi)$100 Hz.          & pump laser linewidth    \\
      $\mu$		& 3.8 mHz.          & $1/2\times$Bril. ampl. rate per photon    \\
      \hline
    \end{tabular}
  \end{center}
\end{table}
 
\section{Laser noise dynamics}
In this section we explore the amplitude and phase noise dynamics of cascaded Brillouin lasers. We base this analysis on the effective dynamics described by Eq. \eqref{laser_equation_1}. Consequently our results differ slightly from Loh {\it et al.} \cite{Loh2015} who included non-adiabatic effects of the phonon mode(s) but neglected quantum fluctuations. To explore the phase and amplitude dynamics of a cascaded order Brillouin laser, we explicitly solve Eq. \eqref{laser_equation_1}, linearized for small $\delta \alpha_m$.  To obtain the desired equations of motion, we combine Eq. \eqref{laser-ansatz} with Eq. \eqref{laser_equation_1}, keep terms to linear order in $\delta \alpha_m$, take the real and imaginary parts, and use the relations between the steady-state amplitudes to find  
\begin{align}
\label{}
   \dot{\delta \alpha}_m  = &
  -2(\mu_m' \alpha_{m+1} \delta \alpha_{m+1} - 
  \mu_{m-1}' \alpha_{m-1} \delta \alpha_{m-1})\alpha_m
  \nonumber 
  \\
  & 
   +  {\rm Re}[\tilde{h}_m] - \frac{1}{\alpha_m}\sqrt{\gamma_{\rm ext}}|F_{\rm pump}| \delta_{m0} \delta\alpha_m
   \nonumber 
   \\
  & 
  + \sqrt{\gamma_{\rm ext}} ({\rm Re}[ \tilde{F}_{\rm ext}]-|F_{\rm pump}|)\delta_{m0}
  \\
    \nonumber \\
   \alpha_m \dot{\varphi}_m = &
   -2(\mu_m'' \alpha_{m+1} \delta \alpha_{m+1} + 
  \mu_{m-1}'' \alpha_{m-1} \delta \alpha_{m-1})\alpha_m
   \nonumber \\
 &  
 \quad -\delta\omega_m (1+ 
 \delta\alpha_m/\alpha_m) +  {\rm Im}[ \tilde{h}_m]
 \nonumber 
 \\
  &
  \label{phase-dyn-int}
 \quad  + \sqrt{\gamma_{\rm ext}} {\rm Im}[\tilde{F}_{\rm ext} ]\delta_{m0}
\end{align}
where $\mu_m'' = {\rm Im}[\mu_m]$ 
and $\delta \omega_m = (\mu_m'' \alpha_{m+1}^2 + 
  \mu_{m-1}'' \alpha_{m-1}^2)\alpha_m$.
To obtain the equation above, we have multiplied $F_{\rm pump}$ as well as the Langevin forces by $\exp \{-i\varphi_m\}$, yielding the definitions 
$\tilde{h}_m \equiv {h}_m \exp \{-i\varphi_m\}$ and $\tilde{F}_{\rm ext} \equiv {F}_{\rm ext} \exp \{-i\varphi_m\}$. 

The cascaded Brillouin laser noise dynamics described by the equations above share two important features with typical laser systems. Generally, the laser phase noise and RIN couple, and the dynamics of the various cascaded orders couples. When phase matching is precisely satisfied the laser phase and amplitude decouple. However, the amplitudes of adjacent laser orders continue to interact, resulting complex relaxation oscillation dynamics. 

To begin our discussion of laser noise, we explore the dynamics of the $a_0$ ($m=0$ mode). This mode acts as the pump for the Brillouin laser and has distinct dynamics from those of the other optical modes. Among these distinctions, the $m=0$ mode does not undergo a lasing transition, and it is driven by a noisy external pump laser, noise that is directly transferred to the pump mode and can be fed into cascaded Stokes orders.

\subsection{Pump dynamics}
\begin{figure}
\begin{center}
\includegraphics[width=3.2in]{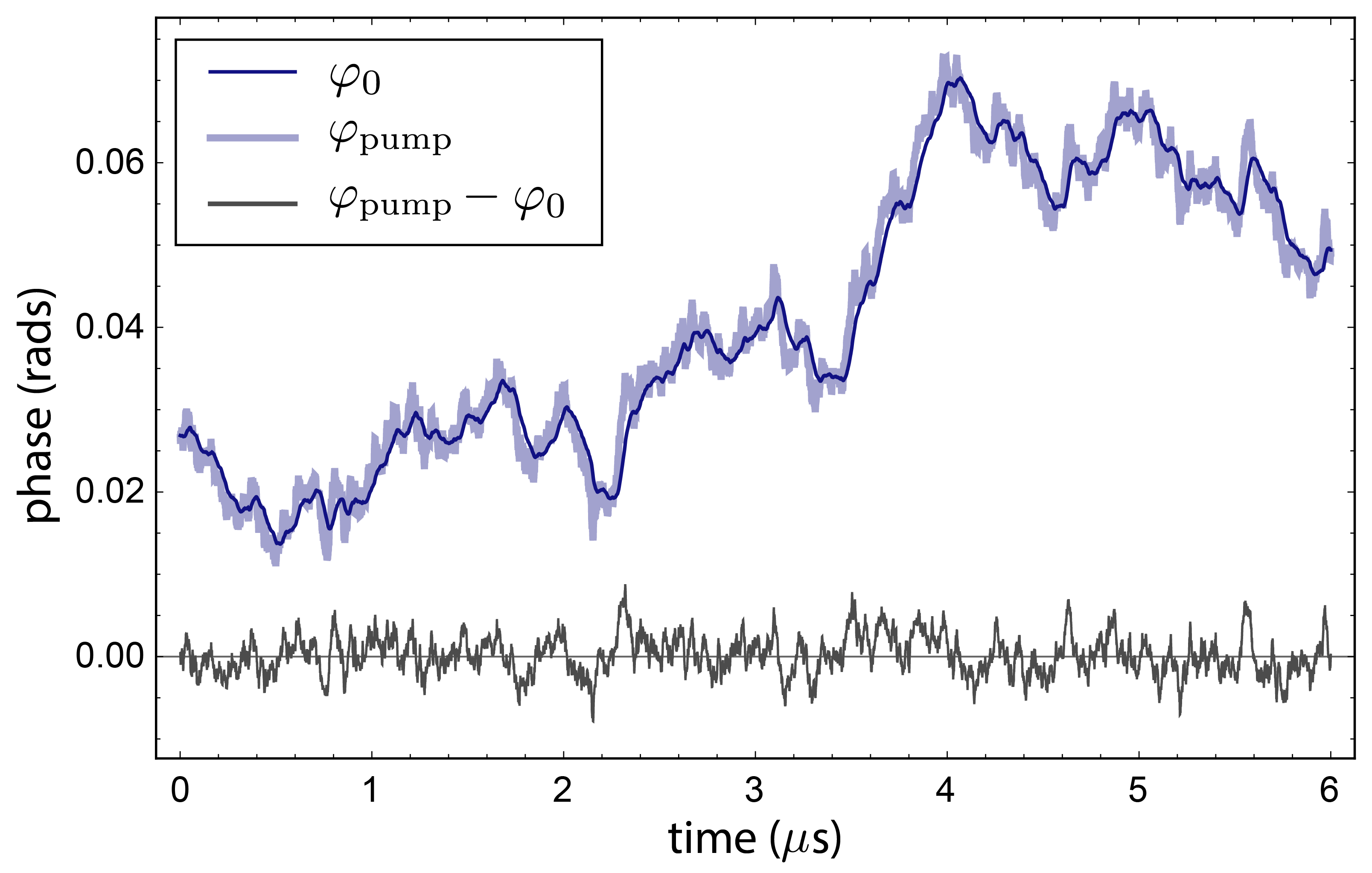}
\caption{Simulated pump phase $\varphi_0$ and external pump phase $\varphi_{\rm pump}$ as a function of time for $P_{\rm pump} = 50 mW$. Simulation parameters given in Tab. \eqref{tab:table1}}
\label{pump-phase}
\end{center}
\end{figure}

In this section we analyze the dynamics $a_0$, the $m= 0$ optical mode, beginning with the phase. To obtain the time-dependence of the pump ($m= 0$ optical mode) we assume that the external pump laser is given by 
\begin{align}
\label{}
F_{\rm pump} = |F_{\rm pump}| 
\exp\{i(\Delta \omega t+ \varphi_{\rm pump})\}
\end{align}
where $\Delta \omega \equiv \omega_{\rm pump} - \omega_0$ is the difference between the external pump laser frequency and the resonance frequency of the pump ($m$=0) mode (see \ref{cascaded-laser-sketch}). As discussed above, we assume the source laser is phase noise dominated, that the amplitude $|F_{\rm pump}|$ is time-independent, and the phase $\varphi_{\rm pump}$ is randomly fluctuating in time with a variance determined by the external pump laser linewidth $\Delta\nu_{\rm pump}$ (see \ref{cascaded-laser-sketch}). We model the behavior of $\varphi_{\rm pump}$ using the phase diffusion model \cite{Debut2000}. These assumptions yield the equation for $\varphi_0$ given by 
\begin{align}
\label{}
  \alpha_0 \dot{\varphi}_0 = & 
   -2\mu_0'' \alpha_{1} \alpha_0 \delta \alpha_{1}   -\delta\omega_0 (1+ 
 \delta\alpha_0/\alpha_0)
  \nonumber \\
 & +  {\rm Im}[ \tilde{h}_0]
  +   \sqrt{\gamma_{\rm ext}} |F_{\rm pump}|
  \sin( \Delta \omega t + \varphi_{\rm pump} - \varphi_0)
\end{align}
where $\delta \omega_0$ is defined just after Eq. \eqref{phase-dyn-int}
To find the dynamics of $\varphi_0$, we assume that
 precise phase matching is satisfied (i.e. the electrostrictive coupling parameter $\mu_m$ is real and $\mu_m'' = 0$), that the pump mode is driven on resonance ($\Delta \omega = 0$), that $ \sqrt{\gamma_{\rm ext}} |F_{\rm pump}| \gg |\tilde{h}_0|$, and that $\tilde{\gamma}_0 \gg 2\pi \Delta\nu_{\rm pump}$. In other words, the last condition means that the optical pumping of our system is not Lorentz limited. 
Under these conditions, the pump phase decouples from the laser amplitudes and adiabatically follows the phase of the external source laser $\varphi_{\rm pump}$, yielding
\begin{equation}
\label{ }
\varphi_0 \approx \varphi_{\rm pump}
\end{equation}
This behavior is illustrated in Fig. \ref{pump-phase} where the time evolution of $\varphi_{\rm pump}$ and $\varphi_{0}$, obtained from stochastic simulations of Eqs. \eqref{Heisenberg-Langevin-1} \& \eqref{Heisenberg-Langevin-2}, is shown. 
Under the same assumptions given above, we find the pump amplitude dynamics given by
\begin{align}
\label{}
 &  \dot{\delta \alpha}_0  \approx 
  -2\mu_0' \alpha_{1} \alpha_0\delta \alpha_{1} 
   - \frac{1}{\alpha_0}\sqrt{\gamma_{\rm ext}} |F_{\rm pump}|\delta\alpha_0
   +  {\rm Re}[\tilde{h}_0]. 
 \end{align}
In the following, we will use these equations, along with equations describing the steady-state powers, to find the RIN and the phase noise in cascaded Brillouin lasers.  
 
\subsection{Cascaded Brillouin laser noise in phase matched systems}
While the development presented up to this point was general, here we restrict our analysis to  perfectly phase matched systems. Under such conditions the laser dynamics dramatically simplifies. We utilize these simplifying assumptions to calculate the laser noise for a range of examples. For the examples considered here, satisfying strict phase matching for all cascaded orders, the coupling parameters are real, i.e. $\mu_m = \mu'_m$ and $\mu''_m = 0$, and the linearized dynamics of the various laser phases decouple from the amplitudes, yielding the laser dynamics described by 
\begin{align}
\label{amp-dyn}
 &  \dot{\delta \alpha}_m  = 
  -2\mu_m \alpha_{m+1} \alpha_m \delta \alpha_{m+1} + 2 \mu_{m-1} \alpha_{m-1} \alpha_m \delta \alpha_{m-1} 
    \nonumber 
  \\
 & \quad \quad \quad \quad  
- \frac{1}{\alpha_0}\sqrt{\gamma_{\rm ext}} |F_{\rm pump}| \delta_{m0} \delta \alpha_{m} 
  +  {\rm Re}[\tilde{h}_m] 
\\
\label{phase-dyn}
 &  \alpha_m \dot{\varphi}_m = 
  {\rm Im}[ \tilde{h}_m] 
\end{align}
where, justified by the dynamics of $\varphi_0$, we have dropped a term proportional ${\rm Re}[ \tilde{F}_{\rm ext}]-|F_{\rm pump}|$. 
Let us take a moment to address a subtlety of the decoupling between the amplitude and the phase dynamics. Recall that the Langevin forces given above are multiplied by factors of the form $\exp\{-i\varphi_m\}$ (see below Eq. \eqref{phase-dyn-int}), and therefore, the amplitude and phase dynamics are coupled, in contrast with the claims above. However, when the correlation time for $h_m$ is short compared to that of the laser phases (which is well-satisfied in typical systems), the correlation properties of $h_m$ and $\tilde{h}_m$ are indistinguishable, and the amplitude and phase dynamics become effectively decoupled. This decoupling enables the laser amplitude and phase noise to be analyzed independently. 

We begin our analysis of laser noise by calculating the RIN. Unlike the laser phases, Eq. \eqref{amp-dyn} shows that the amplitudes of the various laser orders couple together. This coupling can produce relaxation oscillation dynamics with multiple resonant frequencies, depending on the number of lasing modes. Consequently, the RIN must be analyzed case by case.  

\subsubsection{Relative intensity noise}
In this section, we use the decoupled equations Eq. \eqref{amp-dyn} to find the relative intensity noise (RIN) of a cascaded Brillouin laser, quantifying the relative stability of the emitted laser power. For the $m$th laser mode, the RIN $S^{\rm RIN}_m[\omega]$ is defined by the two-sided power spectrum of the relative laser power fluctuations
\begin{equation}
\label{ }
S^{\rm RIN}_m[\omega] = \frac{1}{P_m^2}\int_{-\infty}^\infty d\tau \ e^{i\omega \tau}
\langle \delta P_m(t+\tau) \delta P_m(t) \rangle,
\end{equation}
where $\delta P_m$ represents the time-dependent variation of the laser power from its steady-state value. By using $(P_m +\delta P_m) \propto (\alpha_m + \delta \alpha_m)^2$ and assuming that $|\delta \alpha_m| \ll \alpha_m$, we can express the power spectrum for relative intensity noise in terms of the laser amplitude fluctuations as
\begin{equation}
\label{RIN-formula}
S^{\rm RIN}_m[\omega] =
 \frac{4}{\alpha_m^2}\int_{-\infty}^\infty d\tau  \ e^{i\omega \tau}
 \langle \delta \alpha_m(t+\tau) \delta \alpha_m(t) \rangle. 
\end{equation}
Here, we have neglected subleading terms of order $\delta \alpha^4$. 

In the following, we solve the laser equations for the amplitude dynamics and use Eq. \eqref{RIN-formula} to find explicit expressions for the RIN that depend on the number of cascaded lasing orders. 

\subsubsection{RIN: first order cascading}
We begin by finding the RIN when threshold for the $m=2$ mode has not been met. In this limit, the laser amplitude equations reduce to
\begin{align}
\label{}
 &  \dot{\delta \alpha}_0  = -\frac{1}{\alpha_0} \sqrt{\gamma_{\rm ext}} |F_{\rm pump}| \delta \alpha_0
  -2\mu_0 \alpha_{1} \alpha_0 \delta \alpha_{1} 
  +  {\rm Re}[\tilde{h}_0]
  \\ 
  &  \dot{\delta \alpha}_1  = 
   2 \mu_{0} \alpha_{0} \alpha_1 \delta \alpha_{0}
  +  {\rm Re}[\tilde{h}_1].
\end{align}
As has been described in Ref. \cite{Loh2015}, the amplitude coupling between the pump and first Stokes modes described above leads to relaxation oscillations of energy between the modes, with a frequency given by $\omega^{\rm rel}_{0} \equiv 2 \mu_0 \alpha_0 \alpha_1$  and a damping rate $\Gamma_{\rm RIN} \equiv \sqrt{\gamma_{\rm ext}} |F_{\rm pump}|/\alpha_0$. 

\begin{figure*}
\begin{center}
\includegraphics[width=\textwidth]{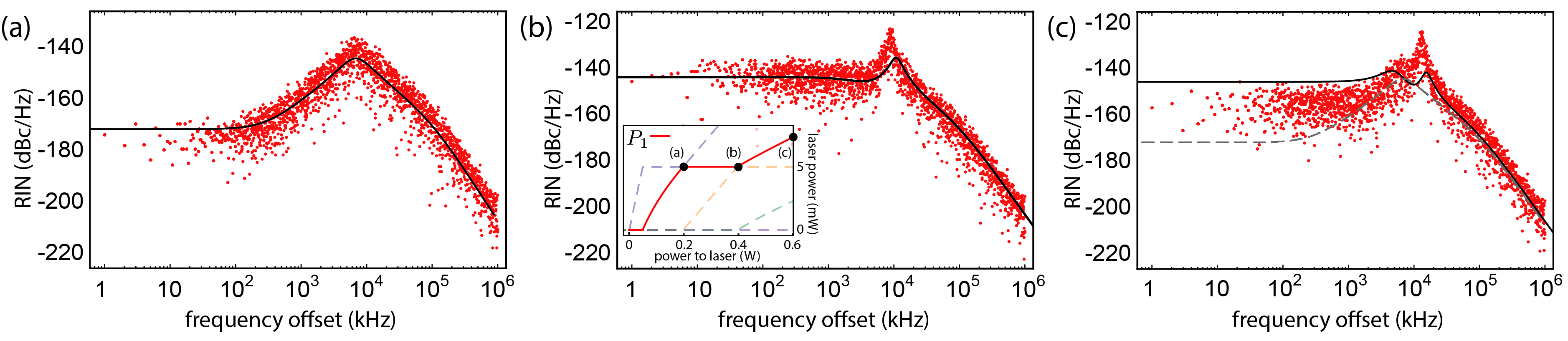}
\caption{Relative intensity noise of the first Stokes order for (a) prior to cascaded lasing (point (a) of inset),
(b) cascaded operation with two laser tone (point (b) of inset), and (c) cascaded operation with 3 laser tones (point (c) of inset). Gray dashed line (c) is the theory curve from (a), included from comparison.
}
\label{RINplot}
\end{center}
\end{figure*}

We solve this coupled set of linear differential equations by Fourier transform, yielding the solution for $\delta \alpha_1$ given by 
\begin{align}
\label{}
\delta \alpha_1(t) = &\int_{-\infty}^\infty \frac{d\omega}{2\pi} \int_{-\infty}^\infty dt_1 \ 
e^{-i\omega(t-t_1)} \chi_{\rm RIN}(\omega) \nonumber \\
&
\times
\bigg( \omega^{\rm rel}_0 {\rm Re}[\tilde{h}_0(t_1)]
+(-i\omega + \Gamma_{\rm RIN}){\rm Re}[\tilde{h}_1(t_1)] 
\bigg)
\end{align} 
where $\chi_{\rm RIN}(\omega) = (-\omega^2 -i \Gamma_{\rm RIN} \omega + {\omega^{\rm rel}_0}^2)^{-1}$. Using the correlation properties for $\tilde{h}_m$ (see Appendix B), we find the two-time correlation function for the amplitude $\langle \delta \alpha_1(t+\tau) \delta \alpha_1(t) \rangle$, the Fourier transform of this correlation function can be used to find the power spectrum for the RIN of a single-mode Brillouin laser, yielding   
\begin{align}
\label{RIN-1}
S^{\rm RIN}_1[\omega] = 
& |\chi_{\rm RIN}(\omega)|^2 \bigg[
\frac{1}{2}{\omega^{\rm rel}_0}^2 \tilde{\gamma}_0  (N_0 +1/2)
\nonumber \\
& +\frac{1}{2}(\omega^2 + \Gamma_{{\rm RIN}}^2) \tilde{\gamma}_1(N_1 +1/2)
\nonumber \\
& +\frac{1}{2} |g_0|^2(n_0 +1/2)
\bigg({\omega^{\rm rel}_0}^2\alpha_1^2 
\nonumber \\
&- 2\omega^{\rm rel}_0 \Gamma_{{\rm RIN}} \alpha_1 \alpha_0
\nonumber \\
& +
(\omega^2 + \Gamma_{{\rm RIN}}^2) \alpha_0^2
\bigg) \frac{\Gamma_0}{\omega^2 + \Gamma_0^2/4}
\bigg].
\end{align} 
Equation \eqref{RIN-1} reproduces the RIN in Brillouin lasers described by Loh {\it et al.} \cite{Loh2015}, when quantum noise is neglected and when $\Gamma_0 \gg \tilde{\gamma}_0$ is assumed. 
In Fig. \ref{RINplot}a, we compare Eq. \eqref{RIN-1} to the RIN power spectrum obtained from stochastic simulations of Eqs. \eqref{Heisenberg-Langevin-1} \& \eqref{Heisenberg-Langevin-2}, both calculations use Tab. \eqref{tab:table1} for input parameters. The agreement between Eq. \eqref{RIN-1} and the laser simulations, shown in Fig. \ref{RINplot}a, justifies the various approximations that led to our analytic expressions describing the RIN. In the next section, we consider RIN in cascaded Brillouin lasers. 

\subsubsection{RIN: Higher order cascading}
Here, we derive general expressions for the RIN for Brillouin lasers that have cascaded to $k$ orders. To formulate this general problem, it is convenient to express the amplitude dynamics in terms of a vector differential equation given by
\begin{equation}
\label{RIN-general}
\dot{\boldsymbol{\delta\alpha}} = - {\bf M} \cdot \boldsymbol{\delta\alpha} + {\rm Re}[\tilde{\bf h}]
\end{equation}
where $\cdot$ denotes matrix multiplication.
Here $\boldsymbol{\delta\alpha}$ and $\tilde{\bf h}$ are column vectors composed of the respective amplitude fluctuations and Langevin forces for each order 
\begin{equation}
\label{ }
\boldsymbol{\delta\alpha} =  
\begin{pmatrix}
       \delta \alpha_k    \\
        \delta \alpha_{k-1} \\
        \vdots \\
        \delta \alpha_1 \\
         \delta \alpha_0 
\end{pmatrix}
\quad \quad \quad
\tilde{\bf h} =  
\begin{pmatrix}
       \tilde{h}_k    \\
         \tilde{h}_{k-1} \\
        \vdots \\
         \tilde{h}_1 \\
          \tilde{h}_0 
\end{pmatrix},
\end{equation}
and the $k\times k$ matrix ${\bf M}$, encoding the amplitude coupling among the various orders, is given by 
\begin{equation}
\label{ }
{\bf M} =  
\begin{pmatrix}
0 &   -\omega^{\rm rel}_{k-1}& \hdots & 	&	        &         &      \\
\omega^{\rm rel}_{k-1}& 0   &  &   	&	        &         &      \\
   \vdots      &    & \ddots &   	&	        &         &    \vdots  \\
&&	     &   0	              &       -\omega^{\rm rel}_{2}       &   0      &   0  \\
&&           &	               \omega^{\rm rel}_{2}  &   0     & -\omega^{\rm rel}_{1}   &  0        \\
&&    	     &                0   &    \omega^{\rm rel}_{1}   &  0     & -\omega^{\rm rel}_{0}     \\
&&         \hdots   &     0   &     0    & \omega^{\rm rel}_{0}    & \Gamma_{\rm RIN} 
\end{pmatrix}
\end{equation}
where $\omega^{\rm rel}_j \equiv 2 \mu_j \alpha_j \alpha_{j+1}$.

This vector differential equation is a compact representation of the dynamics described by Eq. \eqref{amp-dyn}. We obtain the following solution for Eq. \eqref{RIN-general} given by
\begin{align}
\label{RIN-general-soln}
\boldsymbol{\delta\alpha}(t) = \int_{-\infty}^{\infty} \frac{d\omega}{2\pi} 
\int_{-\infty}^{\infty} d t' \ e^{-i\omega(t-t')} {\bf G}[\omega] 
\cdot {\rm Re}[\tilde{\bf h}(t')]
\end{align}
where ${\bf G}[\omega] \equiv [-i\omega {\bf I} + {\bf M}]^{-1}$ (-1 denotes matrix inverse), and ${\bf I}$ is the $k\times k$ identity matrix. The two-sided power spectrum $S^{\rm RIN}_j[\omega]$ for the RIN of the $j$th mode can be obtained by computing the Fourier transform of the two-time amplitude correlation function in Eq. \eqref{RIN-formula}. From the analysis detailed in Appendix C, we find 
\begin{align}
\label{RIN-general-expression}
S_j^{\rm RIN}[\omega] =  \frac{4}{\alpha_j^2}({\bf G}[\omega]\cdot
\overline{\mathcal{C}}[\omega]\cdot
 {\bf G}^\dag[\omega])_{jj} 
\end{align}
where the suffix $jj$ denotes the $jj$ (diagonal)
matrix element of ${{\bf G}[\omega]\cdot
\overline{\mathcal{C}}[\omega]\cdot
 {\bf G}^\dag[\omega]}$, and $\overline{\mathcal{C}}[\omega]$ is a dyadic with matrix elements given by ${\mathcal{C}_{mn}[\omega] = 
 \int_{-\infty}^{\infty} d\tau 
\ e^{i\omega t'} \langle
 {\rm Re}[\tilde{h}_m(t+\tau)]
{\rm Re}[\tilde{h}_n(t+\tau)]
\rangle}$.
To find the RIN for a general case, one finds ${\bf G}[\omega]$ for the relevant number of cascaded orders $k$, and then uses Eq. \eqref{RIN-general-expression}. We give explicit expressions for ${\bf G}[\omega]$ and $\overline{\mathcal{C}}[\omega]$ for a variety cascaded orders in Appendix C. 

In Fig. \ref{RINplot}, we display $S^{\rm RIN}_1[\omega]$, calculated using Tab. \eqref{tab:table1} (and the results of Appendix C), for a range of powers and cascaded orders, the solid black lines are calculated from Eq. \eqref{RIN-general-expression}, and the red dots denote the RIN extracted from simulations of Eqs. \eqref{Heisenberg-Langevin-1} \& \eqref{Heisenberg-Langevin-2}. Figure \eqref{RINplot}(b) shows the RIN for the first Stokes order just prior to threshold for cascading to 3 orders. Although the emitted power for the first Stokes mode is nearly identical for Fig. \ref{RINplot}(a) and Fig. \ref{RINplot}(b), the additional noise channel opened by lasing in the second Stokes order enhances the RIN by nearly 30 dB at low frequencies. As cascading proceeds to higher orders, energy transfer between the cascaded orders produces complex relaxation oscillation dynamics. For example, after third order cascading, the amplitude coupling between the optical modes produces the multipeaked spectra seen in Fig. \ref{RINplot}(c). 

\
\subsubsection{Phase Noise}
Continuing our discussion of noise in cascaded Brillouin lasers, we now explore the phase noise of individual Stokes orders, quantifying the laser frequency stability. Following the conventions of Halford {\it et al.} \cite{Halford1973}, we quantify the phase noise of the $m$th laser order with the power spectrum of phase fluctuations $\mathcal{L}_m(f)$ defined by 
\begin{align}
\label{ell_of_f}
\mathcal{L}_m(f) = \int_{-\infty}^\infty d\tau \ e^{i 2\pi f \tau} 
\langle \varphi_m(t+\tau) \varphi_m(t)
\rangle. 
\end{align}
By integrating Eq. \eqref{phase-dyn} and using the correlation properties of the Langevin force $\tilde{h}_m$ detailed in Appendix B, we find the phase noise given by 
\begin{align}
\mathcal{L}_m(f) \equiv &
\frac{1}{2 \pi f^2} \Delta \nu_m 
\label{phase-noise-1-tone}
\nonumber
\\
 = &\frac{1}{8 \pi^2 \alpha_m^2 f^2} \bigg[ 
  \tilde{\gamma}_m \bigg(\!N_m\!+\!\frac{1}{2}\bigg)
  \nonumber \\
   & \!+\! |g_m|^2 \alpha_{m+1}^2  \bigg(\!n_m\!+\!\frac{1}{2}\bigg)\frac{\Gamma_m}{(2 \pi f)^2\!+\! \Gamma_m^2/4} 
   \nonumber \\
  &  \!+\!  |g_{m-1}|^2 \alpha_{m-1}^2 \bigg(\!n_{m-1}\!+\!\frac{1}{2}\bigg) \frac{\Gamma_{m-1}}{(2\pi f)^2\!+\!\Gamma_{m-1}^2/4} \bigg] 
\end{align}
where, in the top line, we have introduced $\Delta \nu_m$, the fundamental linewidth of the $m$th laser mode.
From left to right, the first term in the brackets originates from thermal and quantum fluctuations of the optical mode, the second term represents the contribution to the phase noise from spontaneous anti-Stokes scattering from of the $m+1$ optical mode, and the last term describes the noise injected into the $m$th mode by spontaneous Stokes scattering from the $m-1$ mode. \

\begin{figure}
\begin{center}
\includegraphics[width=3.4in]{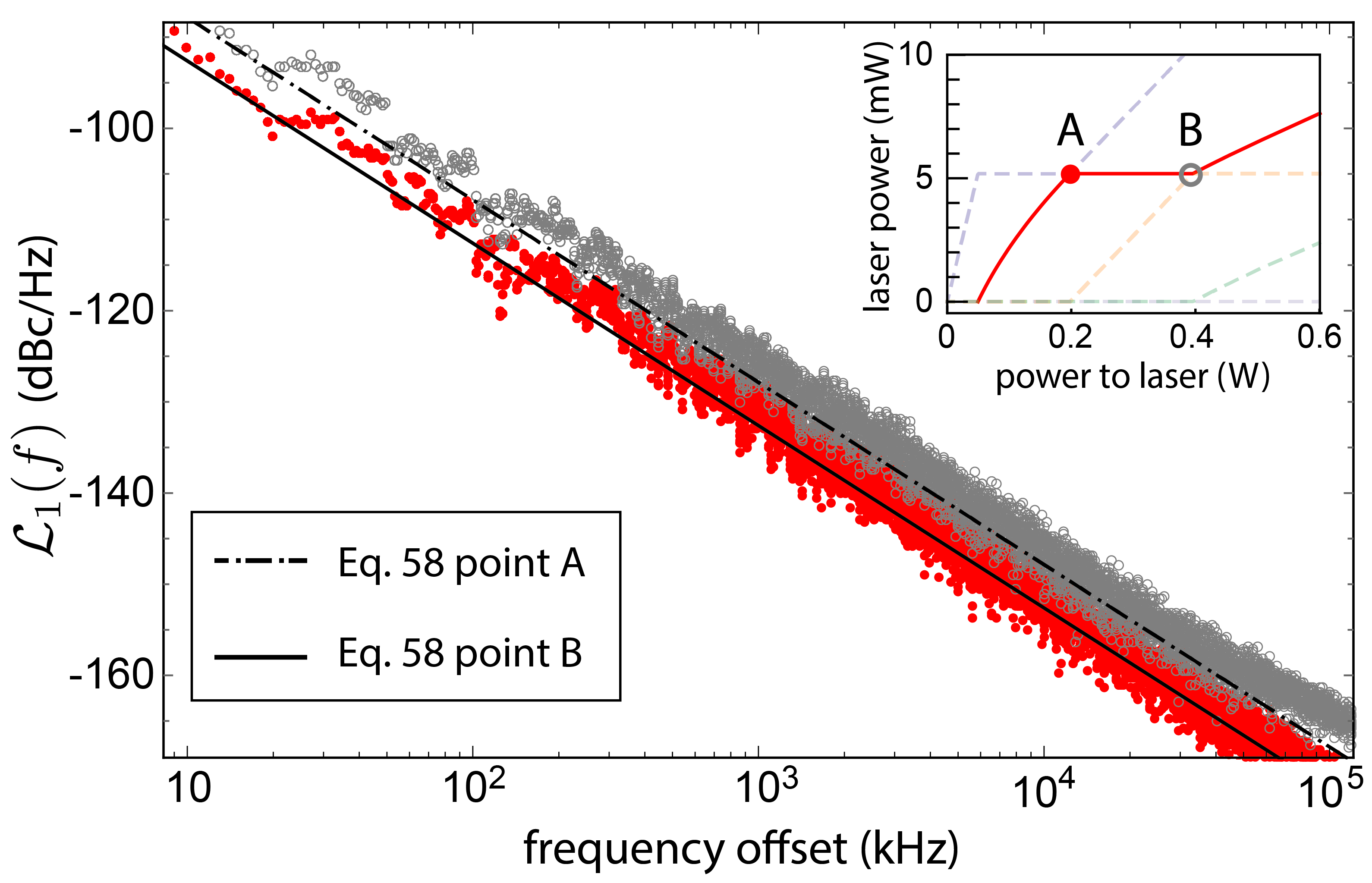}
\caption{Comparison of the phase noise of the first Stokes order, above and below threshold for cascaded lasing. The solid lines are calculated using Eq. \eqref{phase-noise-1-tone}, and the points represent simulated phase noise, open circles for point B of the inset and red points for point A. The emitted laser power is $P_1 = 5.2$ mW for both curves, whereas the power supplied to the laser is respectively $197$ mW and $369$ mW.}
\label{noise-distinction}
\end{center}
\end{figure}

This new result, quantifying the contribution from spontaneous anti-Stokes scattering to the phase noise is one of the central results of this paper. We illustrate the impact of spontaneous anti-Stokes scattering on the phase noise in Fig. \ref{noise-distinction}, showing the phase noise for the first Stokes order below  (point A) and above (point B) threshold for cascaded lasing. Due to power clamping, the emitted power for the first Stokes order at point A and point B is the same, yet, in distinction with insights drawn from the behavior of first order Brillouin lasers, the phase noise is different. The origin of this difference in the phase noise magnitude is due to spontaneous anti-Stokes scattering produced by the second Stokes order. 
\subsubsection{Phase noise of beat notes between cascaded Stokes orders}
In this section, we calculate the coherence properties of microwave signals synthesized using cascaded Brillouin lasers. Cascaded Brillouin lasers offer a compelling method to synthesize high-coherence microwaves \cite{Li2013}. During cascaded operation, a Brillouin laser can co-emit a number of high-coherence laser tones that are spaced by $\sim 10$s of GHz in frequency. By photomixing this laser emission on a high-speed receiver, a coherent electrical signal is produced at the beat frequencies of the various Stokes orders. 

To quantify the coherence of microwave signals synthesized using cascaded Brillouin lasers, we calculate the phase noise power spectrum of the beat note between two distinct Stokes orders. We represent such a beatnote $\beta_{mm'}$, between the $m$th and $m'$th modes, by
\begin{equation}
\label{ }
\beta_{mm'} = a_m^\dag a_{m'} = (\alpha_m+\delta \alpha_m)(\alpha_{m'}+\delta \alpha_{m'})e^{-i\varphi_m}e^{i\varphi_{m'}}.
\end{equation}
If $m \neq m'\pm 1$ then the exponents can be combined (i.e. these phases commute as quantum operators) to yield 
\begin{align}
\label{}
    e^{-i\varphi_m}e^{i\varphi_{m'}} = e^{-i(\varphi_m-\varphi_{m'})}
\end{align}
which gives the beat note phase $\Delta \varphi_{mm'} \equiv \varphi_m-\varphi_{m'}$. 

We calculate the $m$-$m'$ beatnote phase noise power spectrum  $\mathcal{L}_{m,m'} (f)$ by taking the Fourier transform of the two-time beatnote phase correlation function. Given that the phases of the two cascaded laser orders are uncorrelated, the beat note phase correlation function is given by the sum of the phase correlation functions of the individual orders
\begin{align}
\label{}
   \langle \Delta \varphi_{mm'}(t+\tau) \Delta \varphi_{mm'}(t) \rangle 
= & \langle \varphi_{m}(t+\tau)  \varphi_{m}(t) \rangle
\nonumber \\
& \quad \quad
+\langle \varphi_{m'}(t+\tau)  \varphi_{m'}(t) \rangle, 
\end{align} 
yielding the power spectrum for the beat note phase given by
\begin{align}
\label{bn-pn}
\mathcal{L}_{m,m'} (f) = & \mathcal{L}_{m} (f)+\mathcal{L}_{m'} (f).
 \end{align}
This result, combined with Eq. \eqref{phase-noise-1-tone}, shows us that the linewidth of the beatnote $\Delta \nu_{m,m'}$ is an upper bound on the linewidths of the individual tones (i.e. $\Delta \nu_{m,m'} \geq \Delta \nu_m, \Delta \nu_{m,m'} \geq \Delta \nu_{m'})$. 
In the low frequency limit, i.e. $2\pi f\ll \Gamma_m, \Gamma_{m'}$, and by using the recursion relations for the power (Eq. \eqref{recursion}), $\mathcal{L}_{m,m'} (f)$ becomes
\begin{align}
\label{}
\mathcal{L}_{m,m'} (f) \approx 
&   \sum_{j=m,m'}\frac{1}{8 \pi^2 \alpha_j^2 f^2} \bigg[  \tilde{\gamma}_j (N_j+n_{j-1}+1)
\nonumber \\
&   \quad \quad \quad \quad  + 2 \mu_j \alpha_{j+1}^2  (n_j+n_{j-1}+1)\bigg].
 \end{align}
This expression, quantifying the phase noise of beat notes between distinct cascaded laser orders, is the second major result of this paper.  

As a concrete example, we give the phase noise of the beat note between the first and third Stokes orders for a laser that has cascaded to 3 orders. Assuming that the Brillouin coupling and optical decay rates for the 1st and 3rd modes are the same, we find
 \begin{align}
\label{beatnote-PSD}
\mathcal{L}_{1,3} (f)   \approx & \frac{\tilde{\gamma}}{8 \pi^2 f^2}   
\bigg[ \frac{1}{ \alpha_1^2}(N_1\!+\!2 n_{0}\!+\!2\! +\! n_1
)  
\!+\!
\frac{1}{\alpha_3^2}(N_3\!+\!n_{2}\!+\!1)
  \bigg]
   \end{align}
where the clamped value for $\alpha_2$ has been used.
Note that because $\alpha_1$ and $\alpha_3$ are connected by the recursion relation Eq. \eqref{recursion}, a measurement of the emitted power of either order is sufficient to predict the phase noise of the beat note.

\begin{figure}
\begin{center}
\includegraphics[width=3.4in]{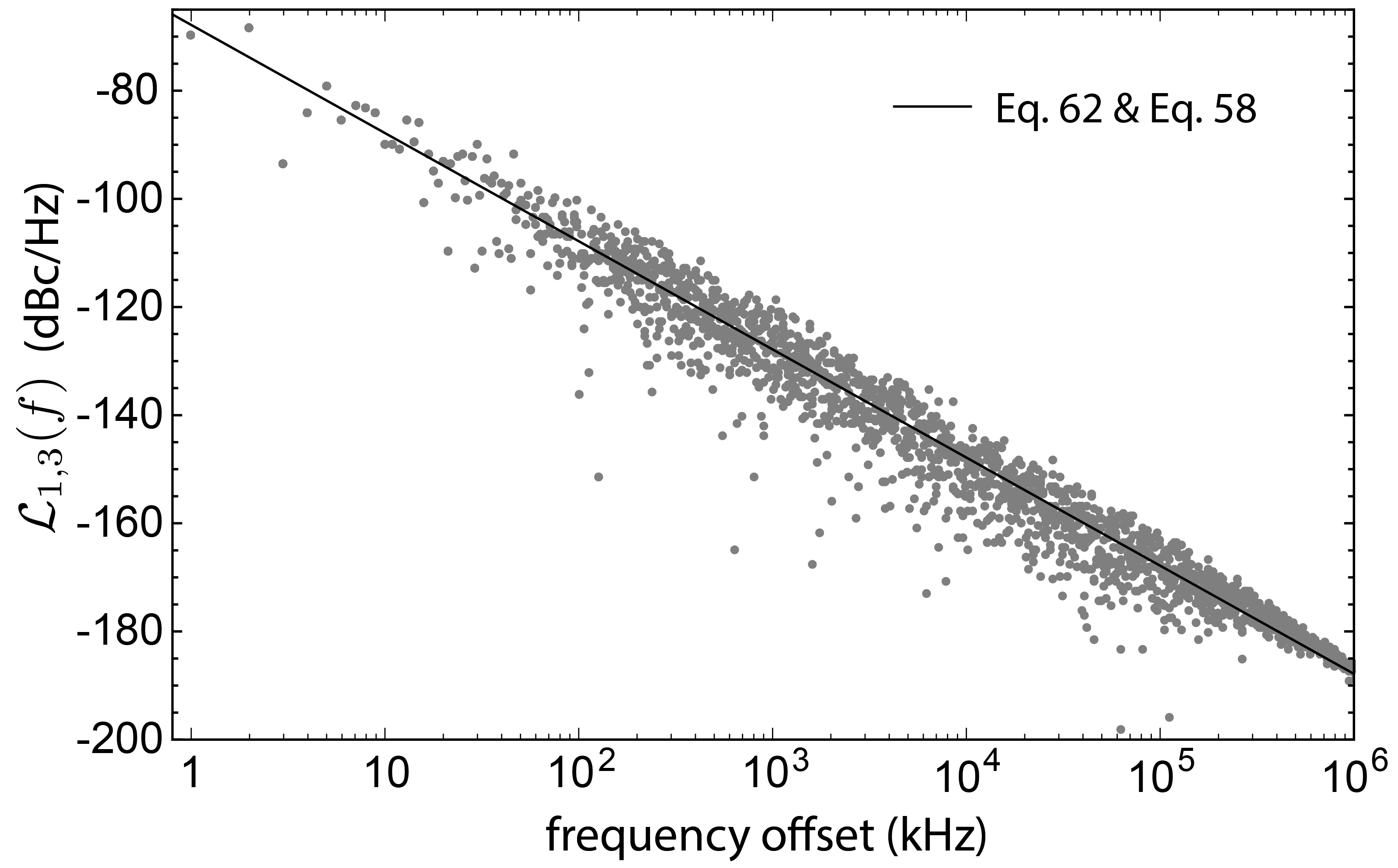}
\caption{Phase noise of beat note of the first and third Stokes laser orders for parameters given in Tab. \eqref{Parameter-Table}. The power spectrum given by Eq. \eqref{beatnote-PSD}, is represented as the solid line. The gray dots represent the simulated beat note phase noise power spectrum obtained by numerically solving Eqs. \eqref{laser_equation_1}. The on-chip pump power is 756 mW.}
\label{beatNotePN}
\end{center}
\end{figure}

Figure \eqref{beatNotePN} compares Eq. \eqref{beatnote-PSD} to the beatnote phase noise obtained by simulating Eqs. \eqref{Heisenberg-Langevin-1} \& \eqref{Heisenberg-Langevin-2}, showing excellent agreement between the theory and simulation. 

As another key result of this paper, Eq. \eqref{beatnote-PSD}
enables the linewidth of the individual optical tones to be quantified by measuring the phase noise of the beatnote and the relative powers of the relevant emitted orders. This analysis can be done by using the theoretical form for the laser linewidths. For a Brillouin laser cascaded to 3 orders and having equal optical decay rates and Brillouin couplings, Eq. \eqref{phase-noise-1-tone} gives the linewidths of the first and third Stokes orders as
\begin{align}
\label{}
    \Delta\nu_1  &  = \frac{\tilde{\gamma}}{4 \pi \alpha_1^2}   
(N_1+2 n_{0}+2 + n_1)  
 \\
  \Delta\nu_3  &  =
 \frac{\tilde{\gamma}}{4 \pi \alpha_3^2}
(N_3+n_{2}+1). 
\end{align}
Using these relations, we find 
\begin{align}
\label{}
 \Delta\nu_3  &  =
\frac{(N_3+n_{2}+1)}{(N_1+2 n_{0}+2 + n_1)}
\frac{\alpha_1^2}{\alpha_3^2} \Delta \nu_1.
\end{align}
For a Brillouin laser operating at room temperature, the thermal occupation of the optical modes is much less than one, while the phonon modes are highly excited $n_0\approx n_1 \approx n_2  \gg 1$. For these conditions, we find $\Delta\nu_3   \approx
P_1/(3 P_3) \Delta \nu_1$
which yields the following relationship between the beatnote linewidth and the linewidth of the first Stokes for the specific example considered here
\begin{equation}
\label{bn_lw}
\Delta\nu_{1,3} \approx [1+P_1/(3P_3)] \Delta \nu_1.
\end{equation}

Equation \eqref{bn_lw} is of paramount importance in designing and characterizing cascaded-order Brillouin (or Raman) lasers. This expression provides a method to assess the Stokes order linewidths without knowing values that have high measurement uncertainty (e.g., fiber coupled power, resonator coupling and ringdown under thermal variations and non-cold cavity conditions).
Consequently, an independent measurement of the beat note phase noise and the relative emitted optical powers of the lasing orders can be used to determine the optical linewidths. 

\subsubsection{Linewidth power dynamics in cascaded Brillouin lasers}
Thus far, we have shown how the noise depends upon the powers of each of the laser orders. Here, we combine these noise results with the steady state power dynamics of Sec. IV to describe the evolution of the laser linewidth with the power supplied to the laser.

\begin{figure}
\begin{center}
\includegraphics[width=3.4in]{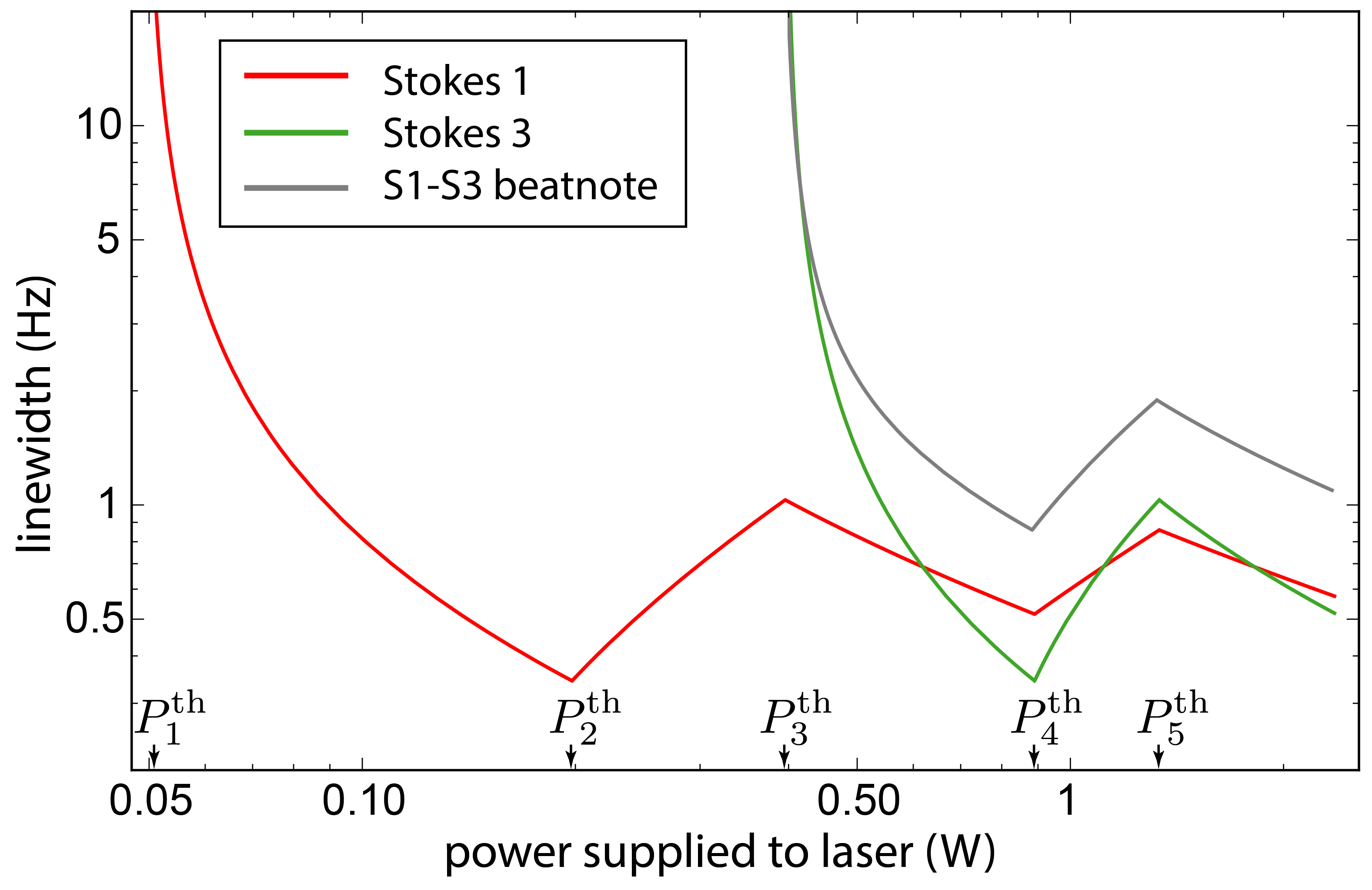}
\caption{Power dynamics of the first Stokes (red), third Stokes (green), and first-third beatnote (gray) linewidth.}
\label{phaseNoisePowerDynamics}
\end{center}
\end{figure}

Figure \eqref{phaseNoisePowerDynamics} shows the evolution of the linewidth of the first and third Stokes orders as well as their beatnote as a function of power supplied to the laser. Beginning at powers below threshold for the second Stokes order, the linewidth $\Delta \nu_1$ (blue curve) decreases inversely with the emitted power $P_1$. Once the threshold power is met for the second Stokers order $P^{\rm th}_2$, Fig. \ref{phaseNoisePowerDynamics} shows a sharp rise in $\Delta \nu_1$. This rise is due to the excess noise injected into the first Stokes order by spontaneous anti-Stokes scattering from the $m=2$ mode. The third Stokes order and the beatnote display similar behavior to that of the first order. Figure \eqref{phaseNoisePowerDynamics} shows that the linewidth of the beatnote and various laser tones exhibits highly nontrivial power dynamics, and, in contrast with single-mode Brillouin lasers, intermediate powers may be preferable for the purpose of producing highly coherent optical or microwave signals.  
 
\section{Conclusion}
In this paper, we explored the power and noise dynamics of cascaded Brillouin lasers. We based this exploration on analytical and numerical studies of a coupled-mode laser model that captures the critical features of cascaded Brillouin lasers. To streamline our theoretical analyses, we investigated the physics of this model under the following simplifying conditions, (1) that the temporal decay rate of the acoustic fields is much faster than the optical fields, and (2) that phase matching is satisfied for all optical fields participating in cascaded lasing. 
Under these conditions, we showed that the laser dynamics can be described by a set of nonlinear stochastic differential equations driven by colored multiplicative noise, and when linearized for small fluctuations around steady-state, the amplitude and phase dynamics decouple. 
Utilizing this drastic simplification, we found the steady-state laser power, and the phase and amplitude dynamics under a variety lasing conditions, yielding the laser RIN and phase noise, as well as the phase noise of beatnotes between distinct laser orders. To corroborate these analytical calculations, we performed stochastic simulations of the full laser model (without the assumptions listed above), and compared the output of these simulations to our theoretical results.

We showed that the dynamics of cascaded lasers contrasts with single-mode Brillouin lasers. This contrasting behavior originates from new noise channels opened by cascaded lasing. We demonstrated that these new noise channels can dramatically enhance the noise of a given laser order (e.g. see Fig. \ref{RINplot}), modifying the laser linewidth and enhancing the RIN. 

We have also presented a simple method to extract Stokes order linewidths, by knowing only the microwave beat-note phase noise and relative optical Stokes order powers. This technique will prove invaluable for assessing sub-Hz linewidths, and optimizing performance for optical and microwave applications.

In the future, we anticipate that these results will provide a valuable toolset to assess the performance or sensitivity of applications of cascaded Brillouin lasers ranging from optical gyroscopes to coherent microwave generation.  

\appendix

\section{Optomechanical coupling rate}
The coupling rate $g_m$ is quantified by the spatial overlap of the acoustic and optical modes that participate in Brillouin scattering. For a laser based on backward Brillouin scattering in a ring resonator, the coupling rate is given approximately by 
\begin{align}
\label{}
    g_m \approx - i \sqrt{\frac{\hbar \omega_m \omega_{m+1}}{
    2 \rho \Omega_m L}} \frac{\omega_m}{v_{p,m}}(\varepsilon-1) \int_{\rm WG} \!\!\!\! d^2 x \ \mathcal{E}_m \mathcal{E}^*_{m+1} \mathcal{U}_m
\end{align}
where $\mathcal{E}_m$ and $\mathcal{U}_m$ are respective mode profiles of the optical and acoustic fields, $\varepsilon$ is the relative permittivity of the waveguide material, and the suffix WG denotes integration across the waveguide cross section. These profiles are normalized across the resonator waveguide cross section, so that 
\begin{align}
\label{}
\int_{\rm WG} \!\!\!\! d^2 x \ |\mathcal{E}_m|^2 = 1
\quad \quad 
 \int_{\rm WG} \!\!\!\! d^2 x \ |\mathcal{U}_m| = 1.
\end{align}
To calculate the coupling rate above, we have treated the optical resonator as a linear waveguide with periodic boundary conditions along the propagation direction. This expression must be generalized to describe resonators with a radius of curvature that is comparable to the mode field diameter. In such systems, the phase accumulated by the inner and outer extreme of the optical mode envelope can be significantly different, leading to the failure of the approximation described above.

\
\section{Correlation functions for the phonon fields and the Langevin forces}
In this section, we evaluate the correlation properties of all of the Langevin forces that are required to calculate the laser noise. First, we begin by finding the two-time correlation function for $\hat{b}_m$.
\subsection{Two-time phonon correlation functions}
Using the solution for $\hat{b}_m$ given in Eq. \eqref{b_Langevin} and the properties $\xi_m$, we find the two-time phonon correlation functions given by 
\begin{align}
\langle \hat{b}^\dag_m(t) \hat{b}_{m'}(t') \rangle = & \delta_{mm'} n_m e^{-\frac{\Gamma_m}{2}|t-t'|}
\\
\langle \hat{b}_m(t) \hat{b}^\dag_{m'}(t') \rangle = & \delta_{mm'} (n_m+1) e^{-\frac{\Gamma_m}{2}|t-t'|},
\end{align}
giving the appropriate equal time expectation value the phonon number operator and preserving the commutation relations for phonon annihilation and creation operators. 
\subsection{Correlation properties of the Langevin force $\tilde{h}_m$}
By using the correlation properties described in Eq. \eqref{Langevin-corr} along with the correlation properties of $\hat{b}_m$, we find
\begin{widetext} 
\begin{align}
\label{h-correlator}
  \langle \tilde{h}_m(t) \tilde{h}^\dag_{m'}(t') \rangle = & 
   \langle [ \eta_{m}(t) - i g_m \alpha_{m+1} e^{i\varphi_{m+1}(t)}  \hat{b}_m(t) - i g^*_{m-1}\hat{b}^\dag_{m-1}(t) \alpha_{m-1}e^{i\varphi_{m-1}(t)}]
   \nonumber \\
   & \times [ \eta^\dag_{m'}(t') + i g^*_{m'} \alpha_{m'+1} e^{-i\varphi_{m'+1}(t')} \hat{b}^\dag_{m'}(t') + i g_{m'-1}\hat{b}_{m'-1}(t') \alpha_{m'-1} e^{-i\varphi_{m'-1}(t')} ] 
   \rangle \nonumber
   \\
   = & \langle  \eta_{m}(t) \eta^\dag_{m'}(t')\rangle
     + |g_m|^2 \alpha_{m+1}^2  \langle \hat{b}_m(t)  \hat{b}^\dag_{m'}(t') \rangle
     \langle e^{i\varphi_{m+1}(t)} e^{-i\varphi_{m'+1}(t')} \rangle
    \nonumber \\
    & +  |g_{m-1}|^2 \alpha_{m-1}^2 \langle \hat{b}^\dag_{m-1}(t) \hat{b}_{m'-1}(t')
     \rangle  \langle e^{i\varphi_{m-1}(t)} e^{-i\varphi_{m'-1}(t')} \rangle
      \nonumber
     \\
     \approx & \delta_{mm'}\bigg[ \tilde{\gamma}_m (N_m+1)\delta(t-t')
     + |g_m|^2 \alpha_{m+1}^2  (n_m+1) e^{- \frac{\Gamma_m}{2}|t-t'|}
    +  |g_{m-1}|^2 \alpha_{m-1}^2 n_{m-1} e^{- \frac{\Gamma_{m-1}}{2}|t-t'|} \bigg].
\end{align}
In the last line we have assumed that the correlation time for the phases is long compared to the phonons. To understand this approximation, assume that the phase noise is a Gaussian process, so that we can evaluate the following expectation value  
\begin{equation}
\label{ }
 \langle e^{i\varphi_{m}(t)} e^{-i\varphi_{m}(t')} \rangle = e^{- \frac{1}{2}\gamma_\phi|t-t'|} 
\end{equation}
where $\gamma_\phi/(2\pi)$ is the linewidth of the phase noise power spectrum. For typical Brillouin lasers $\Gamma_m \gg \gamma_\phi$, allowing $\langle e^{i\varphi_{m}(t)} e^{-i\varphi_{m}(t')} \rangle \to 1$, resulting in the last line of Eq. \eqref{h-correlator}. 

We can obtain $ \langle \tilde{h}^\dag_m(t) \tilde{h}_{m'}(t') \rangle$ by replacing $N_m +1 \to N_m$, $n_m+1 \to n_m$ and $n_{m-1} \to n_{m-1}+1$ in Eq. \eqref{h-correlator}.

In addition, we find 
\begin{align}
\label{}
 \langle \tilde{h}_m(t) \tilde{h}_{m'}(t') \rangle \approx & 
  \langle [ \eta_{m}(t) - i g_m \alpha_{m+1} \hat{b}_m(t) - i g^*_{m-1}\hat{b}^\dag_{m-1}(t) \alpha_{m-1} ][ \eta_{m'}(t') - i g_{m'} \alpha_{m'+1}  \hat{b}_{m'}(t') - i g^*_{m'-1}\hat{b}^\dag_{m'-1}(t') \alpha_{m'-1} ] 
   \rangle \nonumber
   \\
  = &
     -|g_m|^2 \alpha_{m+1} \alpha_{m}  \langle \hat{b}_m(t)  \hat{b}^\dag_{m}(t') \rangle \delta_{m,m'-1}
     -|g_{m-1}|^2 \alpha_{m-1} \alpha_{m}  \langle \hat{b}_{m-1}(t)  \hat{b}^\dag_{m-1}(t') \rangle \delta_{m,m'+1}
        \nonumber
     \\
     = & - \bigg[ 
      |g_m|^2 \alpha_{m+1} \alpha_m (n_m+1) e^{- \frac{\Gamma_m}{2}|t-t'|}\delta_{m,m'-1}
    +  |g_{m-1}|^2 \alpha_{m-1}\alpha_m n_{m-1} e^{- \frac{\Gamma_{m-1}}{2}|t-t'|} 
    \delta_{m,m'+1}\bigg].
\end{align}
A similar calculation yields 
\begin{align}
\label{}
 \langle \tilde{h}^\dag_m(t) \tilde{h}^\dag_{m'}(t') \rangle \approx  & - \bigg[ 
      |g_m|^2 \alpha_{m+1} \alpha_m n_m e^{- \frac{\Gamma_m}{2}|t-t'|}\delta_{m,m'-1}
    +  |g_{m-1}|^2 \alpha_{m-1}\alpha_m (n_{m-1}+1) e^{- \frac{\Gamma_{m-1}}{2}|t-t'|} 
    \delta_{m,m'+1}\bigg].
\end{align}

Using the expressions above, we can find the correlation properties of $ \langle{\rm Re}[ \tilde{h}_m(t)] {\rm Re}[ \tilde{h}^\dag_{m'}(t')] \rangle$,   $\langle {\rm Re}[ \tilde{h}_m(t)] {\rm Im}[ \tilde{h}^\dag_{m'}(t')] \rangle$, and $  \langle {\rm Im}[ \tilde{h}_m(t)] {\rm Im}[ \tilde{h}^\dag_{m'}(t')] \rangle$ which are relevant to  amplitude and phase noise. Defining ${\rm Re}[ \tilde{h}_m(t)] = (\tilde{h}_m(t)+\tilde{h}_m^\dag(t))/2$ and ${\rm Im}[ \tilde{h}_m(t)] = (\tilde{h}_m(t)-\tilde{h}_m^\dag(t))/(2 i)$, we find 

\begin{align}
\label{}
   \langle{\rm Re}[ \tilde{h}_m(t)] {\rm Re}[ \tilde{h}_{m'}(t')] \rangle = \frac{1}{4}\bigg[ 
    \langle \tilde{h}_m(t) \tilde{h}_{m'}(t') \rangle+
     \langle \tilde{h}^\dag_m(t) \tilde{h}_{m'}(t') \rangle+
      \langle \tilde{h}_m(t) \tilde{h}^\dag_{m'}(t') \rangle+
       \langle \tilde{h}^\dag_m(t) \tilde{h}^\dag_{m'}(t') \rangle
   \bigg] 
\end{align}

\begin{align}
\label{ImIm}
   \langle{\rm Im}[ \tilde{h}_m(t)] {\rm Re}[ \tilde{h}_{m'}(t')] \rangle = \frac{1}{4 i}\bigg[ 
    \langle \tilde{h}_m(t) \tilde{h}_{m'}(t') \rangle
    - \langle \tilde{h}^\dag_m(t) \tilde{h}_{m'}(t') \rangle+
      \langle \tilde{h}_m(t) \tilde{h}^\dag_{m'}(t') \rangle
      - \langle \tilde{h}^\dag_m(t) \tilde{h}^\dag_{m'}(t') \rangle
   \bigg] 
\end{align}

\begin{align}
\label{}
   \langle{\rm Im}[ \tilde{h}_m(t)] {\rm Im}[ \tilde{h}_{m'}(t')] \rangle = -\frac{1}{4}\bigg[ 
    \langle \tilde{h}_m(t) \tilde{h}_{m'}(t') \rangle
    - \langle \tilde{h}^\dag_m(t) \tilde{h}_{m'}(t') \rangle-
      \langle \tilde{h}_m(t) \tilde{h}^\dag_{m'}(t') \rangle+
       \langle \tilde{h}^\dag_m(t) \tilde{h}^\dag_{m'}(t') \rangle
   \bigg].
\end{align}

\subsection{Phase noise for Brillouin laser}
In this section we evaluate the correlation function for the phase of an individual laser tone. Representing the solution to Eq. \eqref{phase-dyn} for the phase in Fourier space we find
\begin{equation}
\label{ }
\varphi_m(t) = \lim_{\epsilon \to 0} \frac{1}{\alpha_m} \int_{-\infty}^\infty \frac{d\omega}{2\pi} \int_{-\infty}^\infty d t_1 \frac{e^{-i\omega(t-t_1)}}{-i(\omega+i\epsilon)} {\rm Im}[\tilde{h}_m(t_1)]
\end{equation}
where the parameter $\epsilon$ is included to enforce causality. This expression can be used to find the two-time phase correlation function, giving 
\begin{align}
\label{}
\langle \varphi_m(t+\tau) \varphi_m(t) \rangle  &  =   -\lim_{\epsilon \to 0} \frac{1}{\alpha_m^2} \int_{-\infty}^\infty \frac{d\omega}{2\pi} \int_{-\infty}^\infty d t_1 
 \int_{-\infty}^\infty \frac{d\omega'}{2\pi} \int_{-\infty}^\infty d t_2 \frac{e^{-i\omega(t+\tau-t_1)}e^{-i\omega(t-t_2)}}{(\omega+i\epsilon)(\omega'+i\epsilon)} 
 \langle {\rm Im}[\tilde{h}_m(\tau_1)] {\rm Im}[\tilde{h}_m(\tau_2)] \rangle
 \\
&  =   \lim_{\epsilon \to 0} \frac{1}{\alpha_m^2} \int_{-\infty}^\infty \frac{d\omega}{2\pi} \int_{-\infty}^\infty d \tau' 
 \frac{e^{-i\omega\tau}e^{i\omega \tau'}}{\omega^2+\epsilon^2} 
 \langle {\rm Im}[\tilde{h}_m(\tau_1)] {\rm Im}[\tilde{h}_m(\tau_1-\tau')] \rangle
 \\ 
 & =  \lim_{\epsilon \to 0} \frac{1}{\alpha_m^2} \int_{-\infty}^\infty \frac{d\omega}{2\pi}
\frac{e^{-i\omega\tau}}{2 (\omega^2+\epsilon^2)} \bigg[  \tilde{\gamma}_m (N_m+1/2)
     + |g_m|^2 \alpha_{m+1}^2  (n_m+1/2)\frac{\Gamma_m}{\omega^2 + \Gamma_m^2/4} 
     \nonumber
     \\
     & \quad \quad \quad \quad \quad \quad  \quad \quad \quad \quad \quad \quad  \quad \quad \quad
    +  |g_{m-1}|^2 \alpha_{m-1}^2 (n_{m-1}+1/2) \frac{\Gamma_{m-1}}{\omega^2 + \Gamma_{m-1}^2/4} \bigg]
\end{align}
From this expression, we can read off the phase-noise power spectrum $\mathcal{L}_m(f)$, defined in Eq. \eqref{ell_of_f} \cite{Halford1973}, (where $f$ is $\omega/2\pi$) for the $m$th laser tone
\begin{equation}
\label{}
\mathcal{L}_m(f) = 
\frac{1}{8 \pi^2 \alpha_m^2 f^2} \bigg[  \tilde{\gamma}_m (N_m+1/2)
     + |g_m|^2 \alpha_{m+1}^2  (n_m+1/2)\frac{\Gamma_m}{(2 \pi f)^2 + \Gamma_m^2/4} 
    +  |g_{m-1}|^2 \alpha_{m-1}^2 (n_{m-1}+1/2) \frac{\Gamma_{m-1}}{(2\pi f)^2 + \Gamma_{m-1}^2/4} \bigg].
\end{equation}
This expression can be dramatically simplified in the low-frequency limit, i.e. $2\pi f \ll \Gamma_m$, by using the recursion formula for the steady-state laser powers. In this low-frequency limit, the phase noise reduces to 

 \begin{align}
\label{}
\mathcal{L}_{m} (f) \approx &   
\frac{1}{2 \pi f^2} \underbrace{\frac{1}{4\pi \alpha_m^2} \bigg[  \tilde{\gamma}_m (N_m+n_{m-1}+1)
     + 2 \mu_m \alpha_{m+1}^2  (n_m+n_{m-1}+1) 
  \bigg]}_{\Delta \nu_{m}}
   \end{align}
 which defines the generalized Schawlow-Townes-like linewidth $\Delta \nu_{m}$ for the $m$th order of a cascaded Brillouin laser.

\section{RIN for cascaded Brillouin lasers}
Here, we derive the RIN for a Brillouin laser that has cascaded to $k$ orders. Using Eq. \eqref{RIN-general-soln} we compute the two-time correlation function for $\delta \alpha_j$
\begin{align}
\label{}
\langle \delta \alpha_j(t+\tau) \delta \alpha_{j'}(t) \rangle = 
 \int_{-\infty}^{\infty} \frac{d\omega}{2\pi} 
\int_{-\infty}^{\infty} d t_1 
\int_{-\infty}^{\infty} \frac{d\omega'}{2\pi} 
\int_{-\infty}^{\infty} d t_2 
\ e^{-i\omega(t+\tau-t_1)} 
\ e^{-i\omega'(t-t_2)}
 {\bf G}_{jm}[\omega]
 {\bf G}_{j'n}[\omega']
 \mathcal{C}_{mn}(t_1,t_2)
 \end{align} 
where $\mathcal{C}_{mn}(t_1,t_2)\equiv \langle {\rm Re}[\tilde{h}_m(t_1)] {\rm Re}[\tilde{h}_n(t_2)] \rangle$, ${\bf G}_{jn} = ([-i\omega' {\bf I} + {\bf M}]^{-1})_{jn}$ is the $jn$ matrix element of $[-i\omega' {\bf I} + {\bf M}]^{-1}$, and the Einstein summation convention is used for repeated indices. 

We can simplify this expression by using the properties of $\tilde{h}_m$ given above. These properties show that $\mathcal{C}_{mn}(t_1,t_2)$ is time-stationary, i.e. the correlation function
$\mathcal{C}_{mn}(t_1,t_2) = \mathcal{C}_{mn}(t_1-t_2)$. Using this stationary property, the change of variables given by $t_2 \to t_1 - t'$ can be made, the $t_1$ integral can be done to give $(2\pi)\delta(\omega+\omega')$, allowing the $\omega'$ integral to be done. These steps give
\begin{align}
\langle \delta \alpha_j(t+\tau) \delta \alpha_j(t) \rangle = 
 \int_{-\infty}^{\infty} \frac{d\omega}{2\pi} 
\ e^{-i\omega\tau} 
{\bf G}_{jm}[\omega]
 {\bf G}_{jn}[-\omega]
 \mathcal{C}_{mn}[\omega]
\end{align}
where $\mathcal{C}_{mn}[\omega] = 
 \int_{-\infty}^{\infty} dt' 
\ e^{i\omega t'} \mathcal{C}_{mn}(t')$. 
We obtain $S_j^{\rm RIN}[\omega]$ given in Eq. \eqref{RIN-general-expression} by taking the Fourier transform of $\langle \delta \alpha_j(t+\tau) \delta \alpha_j(t) \rangle$.

Using the results from Appendix B, we find the explicit form for the dyadic matrix $\overline{\mathcal{C}}$ 
\begin{align}
\label{}
\mathcal{C}_{mn}[\omega] = \bigg[\frac{1}{2}\tilde{\gamma}_m(N_m+1/2)
+ \alpha_{m+1}^2 L_m(\omega) + \alpha_{m-1}^2 L_{m-1}(\omega)\bigg]\delta_{mn}
- \alpha_{m+1}\alpha_mL_m \delta_{m,n-1}
- \alpha_{m}\alpha_{m-1}L_{m-1} \delta_{m,n+1}
\end{align}
where $L_m(\omega)$ is defined by 
\begin{align}
\label{}
L_m(\omega) = \frac{1}{4}\mu_m (n_m+1/2)  \frac{\Gamma_m^2}{\omega^2 + \Gamma_m^2/4}.
\end{align}
\subsection{Explicit forms for ${\bf G}$ and $\overline{\mathcal{C}}$}
In matrix form, and by adopting the shorthand 
$\Lambda_j \equiv \frac{1}{2}\tilde{\gamma}_j(N_j+1/2) +\alpha_{j+1}^2 L_{j}+\alpha_{j-1}^2 L_{j-1}$,
we can express $\overline{\mathcal{C}}$ generally as 
\begin{equation}
\label{ }
\overline{\mathcal{C}} =  
\begin{pmatrix}
&\frac{1}{2}\tilde{\gamma}_k(N_k+1/2) +\alpha_{k-1}^2 L_{k-1} &   -\alpha_k \alpha_{k-1} L_{k-1} & \hdots & 	&	        &         &      \\
&-\alpha_k \alpha_{k-1} L_{k-1} & \Lambda_{k-1}   &  &   	&	        &         &      \\
&   \vdots      &    & \ddots &   	&	        &         &    \vdots  \\
&&&	     &   \Lambda_3             &       -\alpha_3 \alpha_2 L_2       &   0      &   0  \\
&&&           &	                -\alpha_3 \alpha_2 L_2  &    \Lambda_{2}     &  -\alpha_2 \alpha_1 L_1   &  0        \\
&&&    	     &                0   &   -\alpha_2 \alpha_1 L_1  &   \Lambda_1  & -\alpha_1 \alpha_0 L_0     \\
&&&         \hdots   &     0   &     0    & -\alpha_1 \alpha_0 L_0  &  \frac{1}{2}\tilde{\gamma}_{0} (N_0+1/2) +\alpha_{1}^2 L_{0}
\end{pmatrix}.
\end{equation}
In contrast, the matrix ${\bf G}$ must be obtained separately for a given number of cascaded laser orders. Below we give the explicit form for ${\bf G}$ for 0,1, and 2 cascaded orders. 
\subsection{1 lasing order, 0 cascaded orders}
\begin{align}
\label{}
 {\bf G}[\omega] = \frac{1}{{\rm det}({i\omega {\bf I} + {\bf M})}}
  \begin{pmatrix}
    -i \omega + \Gamma_{\rm RIN}  &   \omega^{\rm rel}_0 \\
     -\omega^{\rm rel}_0  &   -i \omega
\end{pmatrix}
\end{align}
\begin{align}
\label{}
{\rm det}(i\omega {\bf I} + {\bf M}) = 
-\omega^2 - i \Gamma_{\rm RIN} \omega + {\omega^{\rm rel}_0}^2
\end{align}

\subsection{2 lasing orders, first order cascading}
\begin{align}
\label{}
 {\bf G}[\omega] = \frac{1}{{\rm det}({i\omega {\bf I} + {\bf M})}}
  \begin{pmatrix}
     {\omega_0^{\rm rel}}^2- i \Gamma_{\rm RIN} \omega -\omega^2   &
        \omega^{\rm rel}_1(\Gamma_{\rm RIN}-i\omega)  & 
         \omega^{\rm rel}_0  \omega^{\rm rel}_1
        \\
       -\omega^{\rm rel}_1( \Gamma_{\rm RIN} -i \omega) &
         -i \omega(\Gamma_{\rm RIN} -i \omega) &
         -i   \omega^{\rm rel}_0  \omega
         \\
   \omega^{\rm rel}_0  \omega^{\rm rel}_1   & 
       i \omega^{\rm rel}_0 \omega 		   & 
       {\omega^{\rm rel}_1}^2 - \omega^2 
\end{pmatrix}
\end{align}
\begin{align}
\label{}
{\rm det}(i\omega {\bf I} + {\bf M}) = 
i \omega^3 -  \Gamma_{\rm RIN} \omega^2
-i({\omega^{\rm rel}_0}^2+ {\omega^{\rm rel}_1}^2) \omega + {\omega^{\rm rel}_1}^2 \Gamma_{\rm RIN}
\end{align}
\subsection{Three lasing orders, second order cascading}
\begin{align}
\label{}
 \!\!\!\!\!\!{\bf G}[\omega]\!=\!\frac{1}{X}\! 
  \begin{pmatrix}
   - i  {\omega^{\rm rel}_0}^2\omega\!+\!
 ({\omega^{\rm rel}_1}^2\!-\! \omega^2)(\Gamma_{\rm RIN}\!\!-\!\!i \omega)
 &
 {\omega^{\rm rel}_0}^2{\omega^{\rm rel}_2}
 \!\!\!-\!i \omega {\omega^{\rm rel}_2}(\Gamma_{\rm RIN}\!-\!i\omega)
 &
  {\omega^{\rm rel}_1} {\omega^{\rm rel}_2}(\Gamma_{\rm RIN}\!-\!i\omega)
&
 {\omega^{\rm rel}_0} {\omega^{\rm rel}_1} {\omega^{\rm rel}_2}
\\
-({\omega^{\rm rel}_0}^2{\omega^{\rm rel}_2}
 \!-\!i \omega {\omega^{\rm rel}_2}(\Gamma_{\rm RIN}-i\omega) )  
 &
 - \!i {\omega^{\rm rel}_0}^2 \omega
 \!-\!\omega^2(\Gamma_{\rm RIN}\!-\!i \omega)
 &
 -i\omega {\omega^{\rm rel}_1}(\Gamma_{\rm RIN}-i\omega)
&
- i {\omega^{\rm rel}_0} {\omega^{\rm rel}_1} \omega
\\
 {\omega^{\rm rel}_1} {\omega^{\rm rel}_2}(\Gamma_{\rm RIN}\!-\!i\omega)   
 &
  i\omega {\omega^{\rm rel}_1}(\Gamma_{\rm RIN}-i\omega)
 &
(\Gamma_{\rm RIN}\!-\!i\omega)({\omega^{\rm rel}_2}^2\!\!-\!\omega^2)
&
{\omega^{\rm rel}_0}({\omega^{\rm rel}_2}^2\!\!\!-\!\omega^2)
\\
 - {\omega^{\rm rel}_0} {\omega^{\rm rel}_1}  {\omega^{\rm rel}_2}   
 &
 - i {\omega^{\rm rel}_0}{\omega^{\rm rel}_1}\omega
 &
 -{\omega^{\rm rel}_0}({\omega^{\rm rel}_2}^2\!\!\!-\! \omega^2)
&
\!-i\omega({\omega^{\rm rel}_1}^2\!\!\!\!+\!{\omega^{\rm rel}_2}^2\!\!\!\!\!-\!\omega^2)
\end{pmatrix}
\end{align}
\begin{align}
\label{}
X \equiv {\rm det}(i\omega {\bf I} + {\bf M}) = 
{\omega^{\rm rel}_0}^2{\omega^{\rm rel}_2}^2
-i \Gamma_{\rm RIN} ({\omega^{\rm rel}_1}^2+{\omega^{\rm rel}_2}^2)\omega
-
({\omega^{\rm rel}_0}^2+{\omega^{\rm rel}_1}^2+{\omega^{\rm rel}_2}^2) \omega^2
 + i \Gamma_{\rm RIN} \omega^3
 + \omega^4
\end{align}

 \end{widetext}
 
\bibliography{/Users/rbehunin/Sync/Bibtex/LaserPaper}
\end{document}